\documentclass{aa}
\usepackage{graphicx}
\usepackage{natbib}

\def\rxj{\object{RX~J1856.5$-$3754}}
\def\rxjm{\object{RX~J0720.4$-$3125}}
\let\angstrom\AA
\def\AA{\ifmmode{\hbox{\angstrom}}\else\angstrom\fi}
\def\un#1{\ifmmode{\rm\,#1}\else${\rm\,#1}$\fi}
\let\simgt\ga
\let\simlt\la
\def\Eref#1{Eq.~\ref{eq:#1}}\def\EEref#1{Eqs.~\ref{eq:#1}}
\def\Sref#1{Sect.~\ref{sec:#1}}\def\SSref#1{Sects.~\ref{sec:#1}}
\def\Fref#1{Fig.~\ref{fig:#1}}

\def\d{{\rm d}}
\def\e{{\rm e}}

\begin{document}
\title{An unusual H$\alpha$ nebula around the nearby neutron
       star \rxj\thanks{Based on observations collected at the
European Southern Observatory, Paranal, Chile (ESO Programmes 
63.H-0416 and  65.H-0643)}} 
\titlerunning{An unusual H$\alpha$ nebula around \rxj}
\author{M. H. van Kerkwijk\inst{1} \and S. R. Kulkarni\inst{2}}

\institute{Astronomical Institute, Utrecht University,
           P. O. Box 80000, 3508~TA Utrecht, The Netherlands
     \and  Palomar Observatory, California Institute of
           Technology 105-24, Pasadena, CA 91125, USA}

\offprints{M. H. van Kerkwijk; M.H.vanKerkwijk@astro.uu.nl}

\date{Submitted 2 September 2001}

\abstract{We present spectroscopy and H$\alpha$ imaging of a faint
nebula surrounding the X-ray bright, nearby neutron star \rxj.  The
nebula shows no strong lines other than the Hydrogen Balmer lines and
has a cometary-like morphology, with the apex being approximately
1\arcsec\ ahead of the neutron star, and the tail extending up to at
least 25\arcsec\ behind it.  We find that the current observations can
be satisfactorily accounted for by two different models.  
In the first, the nebula is similar to ``Balmer-dominated'' cometary
nebulae seen around several radio pulsars, and is due to a bow shock
in the ambient gas arising from the supersonic motion of a neutron
star with a relativistic wind.
In this case, the emission arises from shocked ambient gas; we find
that the observations require an ambient neutral Hydrogen number
density $n_{\rm H^0}\simeq0.8\un{cm^{-3}}$ and a rotational energy
loss $\dot{E}\simeq6\times10^{31}\un{erg}\un{s^{-1}}$.
In the second model, the nebula is an ionisation nebula, but of a type
not observed before (though expected to exist), in which the
ionisation and heating are very rapid compared to recombination and
cooling.  
Because of the hard ionising photons, the plasma is heated up to
$\sim\!70000\un{K}$ and the emission is dominated by collisional
excitation.  The cometary morphology arises arises naturally as a
consequence of the lack of emission from the plasma near and behind
the neutron star (which is ionised completely) and of thermal
expansion.  We confirm this using a detailed hydrodynamical
simulation.  We find that to reproduce the observations for this case,
the neutral Hydrogen number density should be $n_{\rm
H^0}\simeq3\un{cm^{-3}}$ and the extreme ultraviolet flux of the
neutron star should be slightly in excess, by a factor $\sim\!1.7$,
over what is expected from a black-body fit to the optical and X-ray
fluxes of the source.  For this case, the rotational energy loss is
less than $2\times10^{32}\un{erg}\un{s^{-1}}$.  
Independent of the model, we find that \rxj\ is not kept hot by
accretion.  If it is young and cooling, the lack of pulsations at
X-ray wavelengths is puzzling.  Using phenomenological arguments, we
suggest that \rxj\ may have a relatively weak, few $10^{11}\un{G}$,
magnetic field.  If so, it would be ironic that the two brightest
nearby neutron stars, \rxj\ and \rxjm, may well represent the extreme
ends of the neutron star magnetic field distribution, one a weak field
neutron star and another a magnetar.
}

\maketitle

\section{Introduction}\label{sec:introduction}

From studies with the {\em EINSTEIN} X-ray satellite and the {\em
ROSAT} all-sky survey, nearly a dozen bright soft X-ray sources have
been identified with nearby neutron stars (\citealt{cbt96,mot00}).
The list includes nearby ordinary pulsars (e.g., \object{PSR
B0656+14}), a millisecond pulsar (\object{PSR J0437$-$4715}) and a
gamma-ray pulsar (Geminga, likely a radio pulsar whose beam does not
intersect our line of sight).  Apart from these, the sample also
includes six radio-quiet, soft X-ray emitting neutron stars whose
origin is mysterious (\citealt{ttz+00}).

The emission from all six sources appears to be completely thermal.
This might make it possible to derive accurate temperatures, surface
gravities, and gravitational redshifts from fits to atmospheric
models.  Therefore, there has been growing interest in the using these
neutron stars as natural laboratories for the physics of dense matter.
In particular, \citet{lp01} show that a radius measurement good to 1
km accuracy is sufficient to usefully constrain the equation of state
of dense matter.  

Of the six sources, the brightest is \rxj\ (\citealt{wwn96}).
\citet{wm97} used the {\em Hubble Space Telescope} ({\em HST}) to
identify \rxj\ with a faint blue star, for which \citet{wal01}
measured a distance of about 60\un{pc} by direct trigonometric
parallax.  \citet{pwl+01} showed that the totality of the data (optical,
UV, EUV and X-ray) can be accounted for by a neutron star with a
surface temperature, $kT_{\rm eff}\simeq50\un{eV}$ and radius
$\sim\!7\,(d/60{\,\rm pc})\,$km.  

As yet, no pulsations have been detected for \rxj\
(\citealt{pwl+01,bzn+01}).  In contrast, for the second brightest
source, \rxjm, \citet{hmb+97}) found clear pulsations, with a period
of 8.39\un{s}.  This source has also been identified with a faint blue
star (\citealt{mh98,kvk98}).  The long rotation period makes the
nature of \rxjm\ rather mysterious.  Models that have received some
considerations are a neutron star accreting from the interstellar
medium and a middle-aged magnetar powered by its decaying magnetic
field (ibid.; \citealt{hk98}).

In order to be able to secure accurate physical parameters from
observations of these sources, we need to understand their nature.  In
particular, we need to know the composition of the surface and the
strength of the magnetic field.  Motivated thus, we have embarked on a
comprehensive observational programme on \rxj.  In our first paper
(\citealt{vkk01}, hereafter Paper~I), we presented the optical spectrum
and optical and ultraviolet photometry from data obtained with the
Very Large Telescope (VLT) and from the {\em HST} archive. We found
that all measurements were remarkably consistent with a Rayleigh-Jeans
tail, and placed severe limits to emission from other mechanisms such
as non-thermal emission from a magnetosphere.

In this paper, the second in our series, we report upon a cometary
nebula around \rxj, which was discovered during the photometric and
spectroscopic observations reported on in Paper I.  We find that the
nebula could be a pulsar bow shock, similar to what is seen around
other pulsars, or a photo-ionisation nebula, of a type that has not
yet been seen anywhere else, but which arises naturally in models of
neutron stars accreting from the interstellar medium
(\citealt{bwm95}).  For either case, the nebula offers a number of new
diagnostics: the density of the ambient gas and the three dimensional
motion of the neutron star, and either a measurement of the rotational
energy loss $\dot E$ from the neutron star, which so far has not been
possible to make for lack of pulsations, or a measurement of the
source's brightness in the extreme ultraviolet, which is impossible to
obtain otherwise because of interstellar extinction.

The organisation of the paper is as follows. First, in \Sref{rxj}, we
briefly summarise the relevant observed properties of \rxj.  Next, in
\Sref{observations}, we present the details of the observations
leading to the discovery of the nebula, and in \SSref{bowshock}
and~\ref{sec:ionisation} we investigate its nature.  

Given the detailed modelling that we have undertaken, the article is
necessarily long.  \Sref{conclusions} offers a summary of our results
and contains a discussion of the clues about the nature of this
enigmatic source offered by the discovery of the nebula.  We recommend
that those of less patient inclination read \Sref{conclusions} first.

\section{\rxj}\label{sec:rxj}

The spectral energy distribution of \rxj\ has been studied in detail
by \citet{pwl+01}, using optical and ultraviolet measurements from
{\em HST} and soft X-ray measurements from {\em EUVE} and {\em ROSAT}.
They find that the overall energy distribution can be reproduced
reasonably well with black-body emission with a temperature
$kT_\infty=48\pm2\un{eV}$,
$n_{\rm{}H}=2.2^{+0.4}_{-0.3}\times10^{20}\un{cm^{-2}}$, and
$R_\infty/d=0.11\pm0.01\un{km}\un{pc^{-1}}$.  The black-body model is
not a very good fit to the X-ray data, which require a somewhat higher
temperature and lower column ($63\pm3\un{eV}$ and
$1.0\pm0.2\times10^{20}\un{cm^2}$, respectively; \citealt{bzn+01}).
This indicates that the real atmosphere is more complicated.  Indeed,
\citet{pwl+01} find that more detailed atmospheric models for
compositions of `Si-ash' and pure Fe give better fits.  For those,
however, one expects strong lines and bands, which appear not to be
present in high-resolution X-ray spectra obtained with {\em Chandra}
(\citealt{bzn+01}).

In Paper~I, we reported VLT photometric and spectroscopic observations
of \rxj. Our VLT spectrum, over the range 4000--7000\,\AA, did not
show any strong emission or absorption features. With considerable
care to photometric calibration, we obtained photometry from VLT and
archival {\em HST} imaging data.  We found that over the entire
optical through ultraviolet range (1500\,\AA--7000\,\AA), the spectral
energy distribution was remarkably well described by a Rayleigh-Jeans
tail, $f_\lambda \propto \lambda^{-4}$ with $A_V=0.12\pm 0.05$ and an
unextincted spectral flux of
$f_{\lambda_0}=(3.36\pm0.17)\times10^{-19}
\un{erg}\un{s^{-1}}\un{cm^{-2}}\un{\AA^{-1}}$ at
$\lambda_0=5000\un{\AA}$.  The reddening corresponds to $N_H\simeq
2.4\times 10^{20}$ cm$^{-2}$ (\citealt{ps95}), consistent with that
estimated from X-ray spectroscopy (see above).

\citet{wal01} used {\em HST} to measure the proper motion and parallax
of \rxj; he found $\mu=332\pm1\un{mas}\un{yr^{-1}}$ at position angle
$\theta=100\fdg3\pm0\fdg1$ and $\pi=16.5\pm2.3\un{mas}$.  The parallax
corresponds to a distance of $61^{+9}_{-8}$\un{pc}.  The proper motion
is directed away from the Galactic plane,
$(\mu_l,\mu_b)=(63.8,-325.8)\un{mas}\un{yr^{-1}}$, and away from the
Upper Scorpius OB association.  \citeauthor{wal01} suggests that \rxj\
may have been the companion of $\zeta$~Oph, which is a runaway star
that likely originated in Upper Sco, but \citet{hdbz01} argue that
PSR~J1932+1059 was the companion to $\zeta$~Oph. In any case, the
proper motion places \rxj\ in the vicinity of Upper Sco association
about a million years ago.  Thus, it is quite plausible that \rxj\ was
born in the Upper Sco association, in which case its age is about a
million years.

The radial velocity of \rxj\ is unknown.  However, assuming an origin
in the Upper Sco association, it should be about
$-55\un{km}\un{s^{-1}}$ (\citealt{wal01}).  If so, the angle between
the plane of the sky and the velocity vector is
$i\simeq60\degr$. However, the radial velocity (and thus $i$) does
depend on the distance of \rxj. For instance, if the distance is 100
pc, then the radial velocity required for an origin in Upper Sco is
about $-20\un{km}\un{s^{-1}}$.

\section{Observations}\label{sec:observations}

In 1999 June, we used the VLT to obtain optical spectra of \rxj.  The
reduction of these data and the resulting spectra are described in
detail in Paper~I.  In the spectra, we discovered extended H$\alpha$
and H$\beta$ emission around the source.  We formed spectral images
along the two slits which we used for the observations -- one roughly
along the proper motion direction and another roughly perpendicular to
it (over stars F and~L, respectively; see \Fref{field} and Fig.~1 in
Paper~I).  From these images, it was clear that the nebula is most
extended along the path that the object has taken.

\subsection{Nebular spectrum}

In \Fref{nebspectrum}, we show the spectrum of the nebula.  This was
formed by extracting a region of 2\arcsec\ around \rxj\ from the
spectral images for both slits, subtracting the contribution from
\rxj, and averaging the two weighted by the exposure time.  The flux
calibration was done with the help of an exposure of the standard star
EG~274 (Paper~I).  From the figure, it is clear that no other lines
than H$\alpha$ and H$\beta$ are observed.  The slightly elevated
continuum at longer wavelengths reflects an increased contribution of
an extended source near the neutron star (see below).

\begin{figure}
\includegraphics[width=88mm]{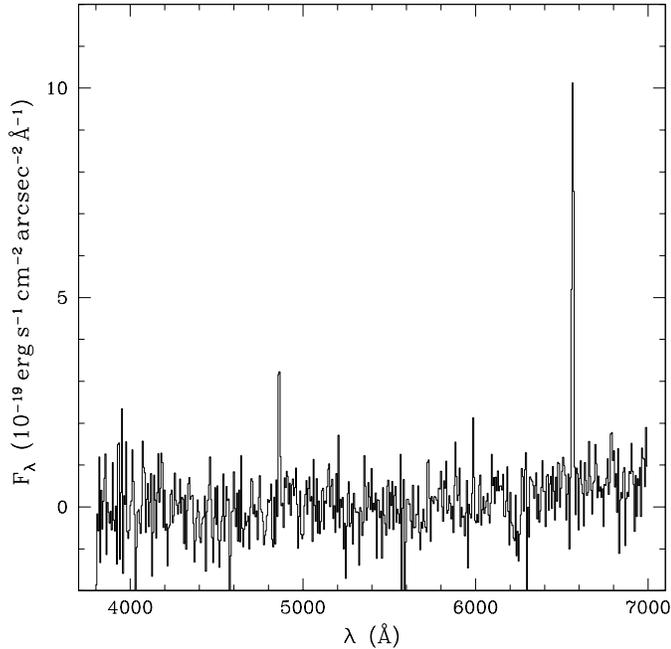}
\caption[]{Spectrum of the nebula around \rxj, showing emission in
H$\alpha$ and H$\beta$, but not in any other lines.  The continuum
emission from \rxj\ has been subtracted.  The slight rise in the
continuum to longer wavelengths is due to a faint, red extended object
close to \rxj\ (see \Sref{Halpha} and \Fref{field}).  }
\label{fig:nebspectrum}
\end{figure}

\subsection{H$\alpha$ image}
\label{sec:Halpha}

To study the nebula in more detail, follow-up H$\alpha$ imaging was
done for us in Service time with the VLT in May 2000.  In addition,
deep B and R-band images were obtained.  The reduction of these images
and the B and R-band photometry are described in Paper~I.  The reduced
H$\alpha$ image is shown in \Fref{field}.  In order to bring out
diffuse emission better, we also show an image in which the stars have
been removed by point--spread function fitting (which was produced in
the process of doing photometry with {\sc daophot}; see below).  Apart
from the faint H$\alpha$ nebula (inside the box), also more extended
diffuse emission is present.  The latter also appears in our B- and
R-band images and hence does not reflect H$\alpha$ emission, but more
likely scattered light, perhaps due to dust associated with the R~CrA
complex.

\begin{figure*}
\includegraphics[width=180mm]{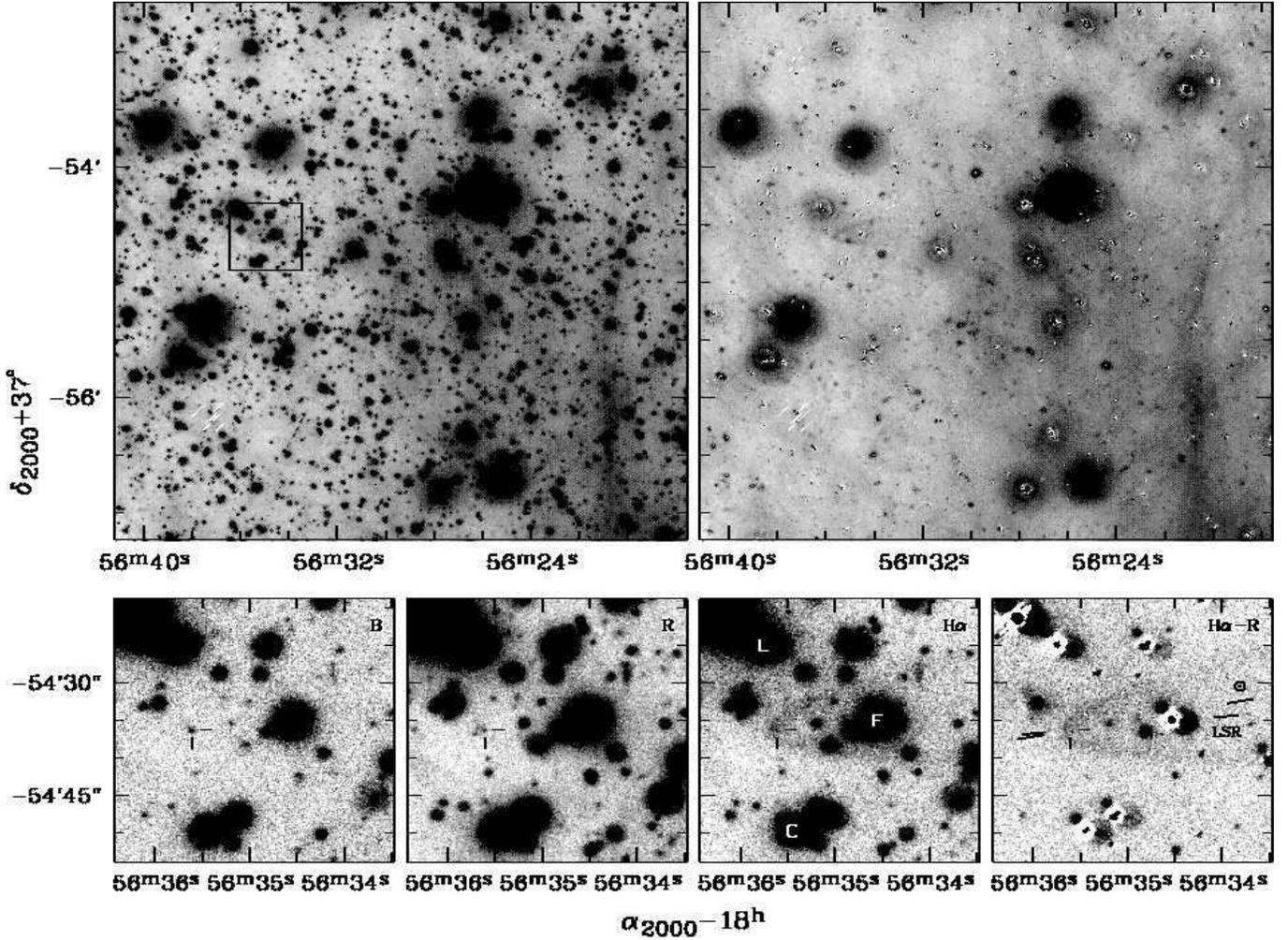}
\caption[]{Images of \rxj, its nebula, and the field.  The upper-left
panel gives an overview of the stacked H$\alpha$ image.  In the
upper-right panel, all but the most overexposed stars have been
removed using point-spread function fitting.  Except for the nebula
(inside the box in the upper-left panel), all diffuse emission that is
seen has matching emission in the B and R images, indicating that it
is due to scattered light.  The lower row of panels shows enlargements
of the area inside the box in the upper-left panel, in B, R, H$\alpha$
and H$\alpha-{}$R.  In the H$\alpha$ image, stars L, C, and F are
indicated.  For the H$\alpha-{}$R image, the R-band image has been
optimally matched to and subtracted from the H$\alpha$ image in order
to remove the contribution from broad-band light.  The tick marks
indicate the position of \rxj, as measured from the B-band image.  In
the R-band image, one sees that there are two objects close to it.
One, about $1\farcs3$ to the ESE, has been detected as well in the
F606W WFPC2 images taken with {\em HST} (\citealt{wal01}; star~115).
The other, about $0\farcs9$ to the NE, is extended and only barely
visible in the {\em HST} images.  These objects give a false
impression of a bright head of the nebula in the H$\alpha$ image; this
impression disappears in the H$\alpha-{}$R image.  In the
H$\alpha-{}$R image, the proper motion is indicated by the longer
dashes, which connect the positions the object had between 60 and 70
years ago and will have in 10 to 20 years from now.  The dash marked
`$\odot$' shows the proper motion as seen from the Sun, while that
marked `LSR' shows the proper motion as seen by an observer moving
with the local standard of rest (and therewith moving with the
interstellar medium local to the neutron star; see
\Sref{orientation}).\label{fig:field}}
\end{figure*}

Figure~\ref{fig:field} also shows enlargements around \rxj.  While the
nebula is clearly visible in the H$\alpha$ image, it is hard to judge
its extent because of the crowded field.  Furthermore, a faint star
and small extended object distort the shape of the nebula's head (the
latter also contributes to the long-wavelength continuum of the
nebular spectrum shown in \Fref{nebspectrum}).  

We tried to remove the contribution of broad-band sources by
subtracting a suitably scaled version of the R-band image.  For this
purpose, we used the image differencing technique of \citet{al98}, in
which the better-seeing image of a pair (in our case, the R-band
image) is optimally matched to the worse-seeing image (H$\alpha$)
before the difference is taken.  The matching is done by constructing
a kernel, convolved with which the PSF of the better-seeing image
matches the PSF of worse-seeing image optimally (in a $\chi^2$ sense).
The resulting H$\alpha-{}$R difference image, shown in \Fref{field},
is not perfect, partly because of differences in the H$\alpha$ to R
flux ratios for different objects, and partly because of internal
reflections in the H$\alpha$ filter, which lead to low-level,
out-of-focus images near every object.  Nevertheless, the nebula can
be seen more clearly, and in particular the confusion near the head is
greatly reduced.  The nebula has a cometary shape and is somewhat
edge-brightened.

\subsection{Orientation}
\label{sec:orientation}

From \Fref{field}, it is clear that the nebula is aligned almost, but
not precisely with the proper motion of \rxj: the position angle of
the nebular axis is $97\degr\pm1\degr$, while the position angle of
the proper motion is $100\fdg3\pm0\fdg1$ (\Sref{rxj}).  This slight
but significant offset is not surprising, because the shape and
orientation of the nebula is determined not by the motion of the
neutron star relative to the Sun, but by its motion relative to the
surrounding medium. 
 
For the ambient interstellar medium, it is reasonable to assume that
its velocity is small relative to the local standard of rest.  If so,
as observed from the Sun, the medium will have an apparent proper
motion $\mu^{\rm ap}$ and radial velocity $v^{\rm ap}_{\rm rad}$, given
by
\begin{eqnarray}
\mu^{\rm ap}_l&=&
  \frac{1}{d}\left[U_\odot\sin l -V_\odot\cos l\right],\nonumber\\
\mu^{\rm ap}_b&=&
  \frac{1}{d}\left[(U_\odot\cos l +V_\odot\sin l)\sin b
                   -W_\odot\cos b\right],\label{eq:reflex}\\
v^{\rm ap}_{\rm rad}&=&
  -(U_\odot\cos l +V_\odot\sin l)\cos b -W_\odot\sin b,\nonumber
\end{eqnarray}
where $(l,b)=(-1\fdg40,-17\fdg22)$ are the Galactic longitude and
latitude of \rxj, and $(U,V,W)_\odot=(10.00,5.25,7.17)\un{km}\un{s^{-1}}$ the
velocity of the Sun relative to the local standard of rest
(\citealt{db98}).  Inserting the numbers, we find $(\mu^{\rm
ap}_l,\mu^{\rm ap}_b)=(-19.3,-36.2)d_{60}^{-1}\un{mas}\un{yr^{-1}}$
and $v^{\rm ap}_{\rm rad}=-7\un{km}\un{s^{-1}}$.

With the above, the proper motion of \rxj\ relative to the nebula, and
therewith, by assumption, relative to the local standard of rest, is
$(\mu^{\rm lsr}_l,\mu^{\rm lsr}_b)=(83.1,-289.7)\un{mas}\un{yr^{-1}}$
for $d=60\un{pc}$.  This implies a total proper motion $\mu^{\rm
lsr}=301.4\,\un{mas}\un{yr^{-1}}$ at position angle $\theta=95\fdg3$;
the latter is close to the angle observed for the nebular axis
(\Fref{field}).  The implied relative spatial velocity on the sky is
$v^{\rm lsr}_\mu=86\un{km}\un{s^{-1}}$, while the relative radial
velocity and total velocity, assuming an origin in the Upper Scorpius
association, are $v^{\rm lsr}_{\rm rad}\simeq-50\un{km}\un{s^{-1}}$
and $v^{\rm lsr}_{\rm tot}=100\un{km}\un{s^{-1}}$.

\subsection{Flux calibration}

No H$\alpha$ calibration images of emission-line objects were taken,
and therefore we used our carefully calibrated spectra of stars~F, L,
and~C (Paper~I) to estimate the throughput.  For this purpose, we
measured instrumental magnitudes on our H$\alpha$ images, using {\sc
daophot} (\citealt{ste87}) in the same way as was done for B and R in
Paper~I, and compared these with the expected magnitudes,
\begin{equation}
m_\alpha = -2.5\log\int A_{\rm UT} R_\alpha(\lambda) R_{\rm open}(\lambda) 
\frac{\lambda}{hc} f_\lambda(\lambda)\,{\rm d}\lambda,
\label{eq:malpha}
\end{equation}
where $A_{\rm UT}=51.2\un{m^2}$ is the telescope area, $R_\alpha$ is
the response of the H$\alpha$ filter, $R_{\rm open}$ the combined
response of all other elements (atmosphere, telescope, collimator,
detector), and $f_\lambda(\lambda)$ the spectrum of the object in
wavelength units.  For $R_\alpha$, we use the response curve from the
ESO web site\rlap{,}\footnote{{\tt
http://www.eso.org/instruments/fors1/Filters/
FORS\_filter\_curves.html}, filter {\tt halpha+59}.} as measured in
the instrument using a grism; the filter has a peak throughput of
$\sim\!65\%$ and full width at half maximum of 60\un{\AA}.  We infer a
total efficiency for all other components of $R_{\rm open}=32\pm1\%$
(assuming $R_{\rm open}$ does not vary strongly over the H$\alpha$
bandpass; we verified that a similar analysis for the R band gives a
consistent result).  Here, the uncertainty includes in quadrature a 2\%
uncertainty in the flux calibration of stars L, C, and~F.  For the
conversion from H$\alpha$ count rates to photon rates, therefore, only
the filter efficiency at H$\alpha$ is required.  We find
$R_\alpha(\lambda_\alpha)=60\pm5\%$, where the relatively large
uncertainty stems from the fact that there is some dependence of the
filter characteristics on position on the detector (even though the
interference filters are located in the convergent beam).

\begin{figure*}
\includegraphics[width=180mm]{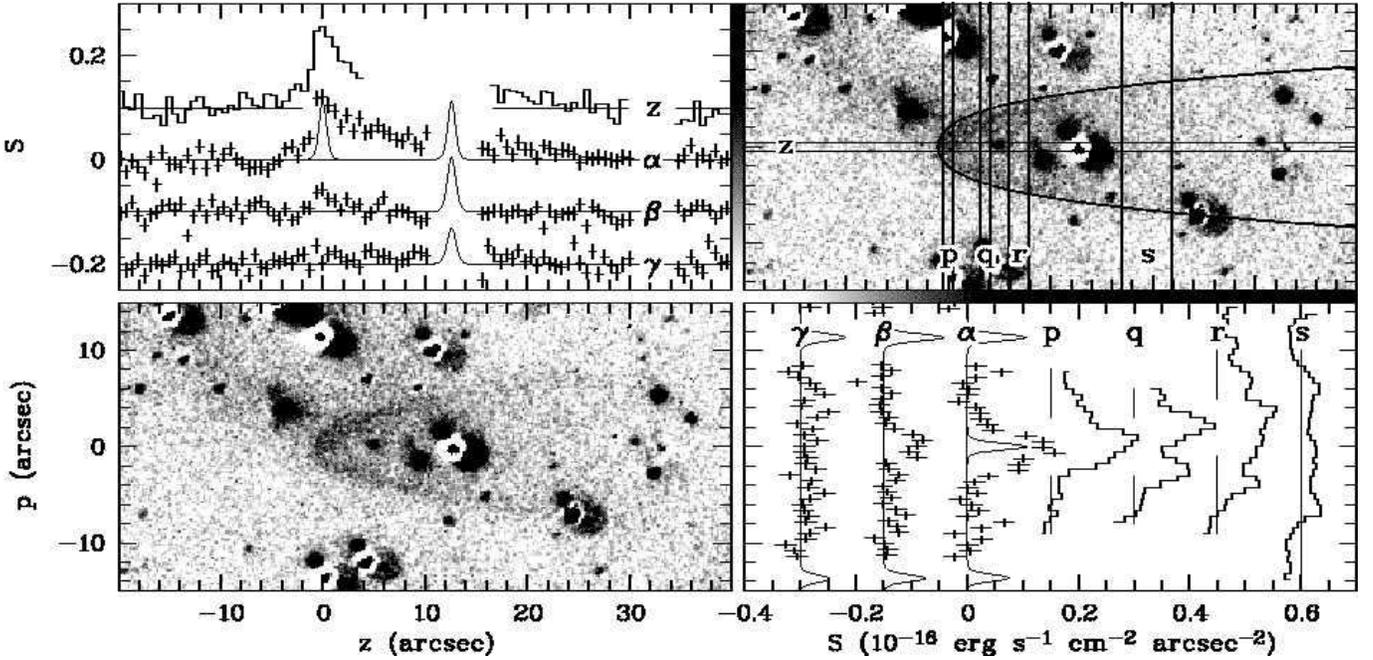}
\caption[]{Flux-calibrated H$\alpha$ images and cross cuts of the
nebula around \rxj.  The {\it lower-left} panel shows the nebula,
using coordinates relative to the neutron star, with the horizontal
axis representing the major axis of symmetry of the nebula.  In the
{\it upper-left} panel, the surface brightnesses along the axis of
symmetry of the nebula are shown.  The points with error bars
represent the H$\alpha$, H$\beta$, and H$\gamma$ brightnesses inferred
from the spectra taken with the slit positioned over star~F, very
close to the axis of symmetry; these are marked as `$\alpha$,'
`$\beta$,' and `$\gamma$,' respectively.  See Fig.~2 and Fig.~1
in Paper~I for star designations including star~F.  The nebula is
faint in H$\beta$ and undetectable in H$\gamma$.  All points are
averages over 3-pixel (0\farcs6) intervals along the slit.  The thin
lines indicate the zero level and show the point-spread function of
star~F (also averaged over 3 pixels).  For H$\alpha$, it also shown at
the position of \rxj.  The histogramme represents H$\alpha$
measurements inferred from the H$\alpha-{}$R image, by averaging over
the numerical slit marked `z' in the upper-right panel; its zero level
is again indicated by a thin line.  The grey scale used in the images
is next to the upper-left and above the lower-right panels.  In the
{\it lower-right} panel, the cross cuts perpendicular to the axis of
symmetry are shown.  The points with error bars represent the
brightnesses inferred from the spectra taken with the slit positioned
over star~L, and the histogrammes those inferred from the image, by
integrating along numerical slits marked `p,' `q,' `r,' and~`s.'  The
zero levels and point-spread functions are indicated by thin lines.
Also displayed in the {\em upper-right} panel is the expected shape
for a bow-shock (see \Sref{bowshock}).
\label{fig:nebula}\label{fig:bs}}
\end{figure*}

\subsection{Results}

In \Fref{nebula}, we show the calibrated H$\alpha-{}$R image, rotated
such that its axis of symmetry is horizontal (assuming a position
angle of $97\degr$, see \Sref{orientation}).  Note that the level of
the background remains somewhat uncertain; indeed, it is the limiting
factor in the resulting H$\alpha$ fluxes.  Also shown in the Figure
are cross-cuts through the image, one along the axis of symmetry, and
four across it at different distances from the head of the nebula.
Furthermore, the H$\alpha$, H$\beta$, and H$\gamma$ surface
brightnesses inferred from the spectra are shown.  The H$\alpha$
brightnesses were derived from the spectral images, by integrating
over the wavelength range of 6554--6570\un{\AA} and subtracting the
average of the integrated fluxes in wavelength ranges 6496--6550 and
6574--6628\un{\AA}.  The H$\beta$ and H$\gamma$ surface brightnesses
were determined similarly.  Note that the H$\alpha$ surface
brightnesses inferred from the spectra are more certain than those
inferred from the H$\alpha$ image, since they suffer much less from
uncertainties in the background level.

The nebula is clearly resolved in all directions.  It reaches
half-maximum brightness $1\farcs0\pm0\farcs2$ in front of the neutron
star, about $2\arcsec$ to each side, and $\sim\!3\arcsec$ behind the
neutron star.  The tail is visible to at least $25\arcsec$ behind the
neutron star.  The peak surface brightness of the nebula is
$1.3\times10^{-17}\un{erg}\un{s^{-1}}\un{cm^{-2}}\un{arcsec^{-2}}$,
corresponding to a photon rate of
$4.3\times10^{-6}\un{s^{-1}}\un{cm^{-2}}\un{arcsec^{-2}}$.

Below we discuss the origin of this nebula and see if we can use it to
shed additional light on the nature of the enigmatic object, \rxj.

\section{A pulsar bow shock?}\label{sec:bowshock}

Cometary nebulae shining in the Balmer lines (with no detectable lines
from any other element) have been seen around a number of pulsars.
The accepted physical model for such nebulae is the pulsar bow-shock
model, which was invoked first by \citet{kh88} to explain a similar
H$\alpha$ nebula around the binary millisecond pulsar \object{PSR
B1957+20}.  Thus, both the similarity of the nebula to nebulae seen
around some pulsars, as well as the reasonable possibility that \rxj\
is a million-year old neutron star (see \Sref{rxj}) and thus might be
a pulsar, motivate us to consider the bow-shock model.

\subsection{Stand-off distance and shape}

Consider a pulsar moving supersonically through partially ionised
interstellar medium of density $\rho_{\rm ism}$.  Along the direction
of motion the medium ram pressure is $\rho_{\rm ism}v_{\rm ns}^2$,
where $v_{\rm ns}$ is the speed of the pulsar.  At the apex of the
nebula, the (quasi) spherical relativistic pulsar wind is balanced by
this ram pressure at a ``stand-off'' distance $r_{\rm a}$, where
\begin{equation}
\frac{\dot E}{4\pi r_{\rm a}^2 c} = \rho_{\rm ism} v_{\rm ns}^2.
\label{eq:bs:standoff}
\end{equation}
Here, $\dot E$ is the luminosity in the relativistic wind, which is
assumed to be primarily in the form of relativistic particles and
magnetic fields.  By considering momentum balance along other
directions, \citet{wil96} derived an analytic solution for the shape
of the nebula,
\begin{equation}
r_\theta\sin\theta = r_{\rm a}\sqrt{3\left(1-\frac{\theta}{\tan\theta}\right)}.
\label{eq:bs:shape}
\end{equation}
Note that in this derivation it is assumed that the shocked gas cools
rapidly, so that thermal pressure is negligible (a two-dimensional
equivalent to the `snow-plough' phase in supernova remnants).  If this
assumption is incorrect then the tail should be slightly wider.

\subsection{Shock emission mechanisms}

Along the cometary surface, one expects two shocks, with in between an
inner region consisting of shocked pulsar wind and an outer region of
shocked interstellar medium.  The former will shine primarily via
synchrotron emission and is best searched for at X-ray (e.g. the
Mouse, \citealt{pk95}) or radio wavelengths (see \citealt{gsf+00} for
a review of radio searches).

All astrophysical shocks appear to be collisionless and mediated via
ionised particles.  Thus, the ambient neutral particles (i.e., atoms)
do not notice the shock, but do find themselves subjected to
collisional excitation and ionisation by shocked electrons and to
charge exchange with shocked protons (\citealt{cr78,ray91}).  Both
these processes can lead to line emission.  Electronic excitation of
the unshocked neutrals will result in a narrow line, reflecting the
velocity dispersion of the unshocked gas.  Charge exchange of a
Hydrogen atom  with a fast-moving proton will result in a Hydrogen atom
having the velocity of the shocked proton and its H$\alpha$ emission
will reflect the large velocity dispersion and the bulk velocity
of the shocked protons (\citealt{cr78,ray91}).  Elements other than
Hydrogen will also be excited collisionally, but the large abundance
of hydrogen and the high temperature ensure that the emitted optical
spectrum is dominated by hydrogen lines.  Hence, the name
``Balmer-dominated'' shocks.

Balmer-dominated shocks are rare because two stringent conditions need
to be met.  First, the ambient ISM has to be (partially) neutral.
Second, the shock has to be ``young.'' The latter condition arises
from the fact that for shock speeds greater than
110\un{km}\un{s^{-1}}, the post-shock gas, if allowed sufficient time
to cool, produces so much ionising flux that the pre-shocked ambient
gas ahead of the shock front becomes almost completely ionised
(\citealt{sm79,mh80}).  Thus, Balmer-dominated shock spectra are
expected only in shocks which have not enough time to sweep a
``cooling column'' density, $N_{\rm cool}=1.8\times
10^{16}v_{100}^{3.6}\un{cm^{-2}}$ (\citealt{hrh87}); here, $v_{100}$
is the shock speed in units of $100\un{km}\un{s^{-1}}$.

Given this reasoning, it might come as a surprise that
Balmer-dominated bow shocks have been found around pulsars with a wide
variety of ages (\citealt{kh88,crl93,bbm+95}).  Pulsar bow shocks are
far from planar, however, and as a result the shocked material
subtends an ever decreasing solid angle as seen from the vantage point
of the stand-off region.  In addition, pulsars spend most of their
time moving through diffuse interstellar medium and thus matter
recombines only far behind the pulsar.  The combination of both
reasons ensures Balmer-dominated pulsar bow shocks can form.

\subsection{Application to \rxj}

Turning now to \rxj, we start by noting that the shape of the nebula
fits reasonably well with the expression given by
Eq.~\ref{eq:bs:shape}, as can be seen in Fig.~\ref{fig:bs}.  By trying
different inclinations $i$ and projected angular stand-off distances
$\theta_\parallel$, we find a best match around $i=60\degr$ and
$\theta_\parallel=1\farcs0$, with a `chi-by-eye' uncertainty of about
$15\degr$ and $0\farcs2$, respectively.  This determination is
consistent with our inference of the radial velocity with respect to
the local medium\ (see \Sref{rxj}).  Note, however, that, as mentioned
above, Eq.~\ref{eq:bs:shape} is derived under the assumption of rapid
radiative cooling; if this is not met, the tail would be wider and
hence the observations would imply inclinations closer to~$90\degr$.
We return to this aspect below.

In the framework of the bow-shock model, apart from the orientation,
there are four unknowns, the relative velocity, the pulsar wind energy
loss $\dot E$, the interstellar medium density $\rho_{\rm ism}$, and
the fraction $\xi_0=n_{\rm H^0}/n_{\rm H}$ of the neutral ambient gas.
In contrast, we have only three measurements: the velocity, the
observed H$\alpha$ flux, $n_\alpha$, and the angular distance between
the neutron star and the apex of the cometary nebula,
$\theta_\parallel$.  The physical length of the standoff distance is
$r_{\rm a}=d\theta_\parallel/\sin{i}=60d_{60}/\sin{i}\un{AU}$, where
$d_{60}$ is the distance in units of 60\un{pc}.  For simplicity, we
will drop the dependence on $\sin i$ in our estimates below, and will
take $v_{\rm ns}=d\mu_{\rm ns}=100d_{60}\un{km}\un{s^{-1}}$.

The H$\alpha$ flux is proportional to the incoming flux of neutral
particles (\ion{H}{i} atoms).  \citet{ray91} states that for a broad
range of shock velocities, about 0.2 H$\alpha$ photons are produced
per neutral particle before that particle is ionised in the post-shock
region.  Unfortunately, there are few calculation for the yield of
H$\alpha$, the number of H$\alpha$ photons per incoming neutral atom,
$f_\alpha$, for low velocity shocks, $\sim\!100\un{km}\un{s^{-1}}$.
According to \cite{cr85}, for a $120\un{km}\un{s^{-1}}$ shock,
$f_\alpha\simeq0.1$ for case~A (i.e., most higher order Lyman photons,
Ly$\beta$, Ly$\gamma$, etc., escape from the shocked region) or
$\sim\!0.23$ for case~B (higher order Lyman photons are absorbed
locally and re-emitted until they are converted; e.g., to H$\alpha$
and two-photon continuum for Ly$\beta$).  We will assume case~A and
normalise the yield accordingly: $f_\alpha=0.1f_{\alpha,0.1}$. 

Considering the head of the nebula only, the expected number
of H$\alpha$ photons is related to the other parameters as follows:
\begin{equation}
n_\alpha 
  = \frac{f_\alpha n_{\rm H^0} \pi r_\perp^2 v_{\rm ns}}{4\pi d^2}
  = \frac{1}{4} f_\alpha n_{\rm H^0} \theta_\perp^2 \mu_{\rm ns} d.
\label{eq:bs:nalpha}
\end{equation}
Here, $r_\perp$ is the radius of the bow shock perpendicular to the
axis of symmetry and $\theta_\perp=r_\perp/d$ the corresponding
angular distance.

With $\theta_\perp=2\arcsec$, the measured $\mu_{\rm ns}$
(\Sref{observations}) and the observed H$\alpha$ photon rate of
$n_\alpha\simeq2\times10^{-5}\un{s^{-1}}\un{cm^{-2}}$, we find
\begin{equation}
n_{\rm H^0} = 0.8\un{cm^{-3}}\; f_{\alpha,0.1}^{-1} d_{60}^{-1}.
\label{eq:bs:n0}
\end{equation}

Noting that the total density of the ambient gas is $\rho_{\rm
ism}=n_{\rm H^0} m_{\rm H}X^{-1}\xi_0^{-1}$, where $X=0.73$ is the
fraction of hydrogen by mass and $\xi_0=n_{\rm H^0}/n_{\rm H}$ is the
neutral fraction, and applying \Eref{bs:standoff}, we find,
\begin{equation}
\dot E \simeq 6\times 10^{31}\un{erg}\un{s^{-1}}\;
              f_{\alpha,0.1}^{-1} {\xi_0}^{-1} d_{60}^{3}.
\label{eq:bs:edot}
\end{equation}
We find that because we have fewer measurements than unknowns, we are
only able to constrain the quantity $\xi_0\dot{E}$.  Since
$\xi_0\leq1$, we obtain a lower limit to $\dot E$.

If, as argued above, \rxj\ is a $t=10^6t_6\un{yr}$ old pulsar, then
assuming that the main energy loss is due to magnetic dipole radiation
(magnetic field strength $B$, braking index $n=3$), we can apply the
usual relations, $\tau_{\rm char}\equiv P/2\dot P=t/\chi$, with
$\chi=1-(P_0/P)^2\simeq1$ (where $P$ and $P_0$ are the current and
birth period, respectively), $\dot{E}=I4\pi^2 \dot P/P^3=I4\pi^2/2\tau
P^2$ (where $I=10^{45}I_{45}\un{g}\un{cm^2}$ is the moment of
inertia), and $B=3.2\times10^{19}(P\dot P)^{1/2}=
3.2\times10^{19}P/(2\tau)^{1/2}\un{G}$, to obtain,
\begin{eqnarray}
P &\simeq& 3.3\un{s}\;\chi^{1/2} I_{45}^{1/2} t_6^{-1/2}
           f_{\alpha,0.1}^{1/2} \xi_0^{1/2} d_{60}^{3/2},
\label{eq:bs:pspin}\\
B &\simeq& 1.3\times10^{13}\un{G}\; \chi I_{45}^{1/2} t_6^{-1} 
           f_{\alpha,0.1}^{1/2} \xi_0^{1/2} d_{60}^{3/2}.
\label{eq:bs:b}
\end{eqnarray}
Since $\chi\leq1$ and $\xi_0\leq1$, the values of $B$ and $P$ are true
upper limits.  The limits on $B$ and $P$ are consistent with those
expected from ordinary million-year old pulsars.  Thus, the bow-shock
model appears to provide a reasonable description of the nebula we
observe.

\subsection{Consistency checks}

We now verify that the bow-shock model is consistent with the
assumptions we made.  First, the shape of the bow shock was assumed to
be specified by Eq.~\ref{eq:bs:shape} and this assumption implicitly
requires that the shock radiate much of the energy. How good is this
assumption?  Every \ion{H}{i} atom comes into the shocked region with
an energy of $\frac{1}{2}m_{\rm H}v_{\rm ns}^2 \simeq 60v_{\rm
ns,100}^2\un{eV}$ where $v_{\rm ns,100}$ is the shock speed in units of
$100\un{km}\un{s^{-1}}$.  Of this, about $\frac{7}{16}$ goes into
shock heating and the remainder into bulk motion.  Only one out of
five hydrogen atom are excited to the $n=3$ level before being
ionised; including other levels, likely the energy loss from the
shocked region is $\sim\!2\times13.6\un{eV}$ per neutral. Thus the
ratio of energy radiated to the incoming kinetic energy is
$\sim\!1.0\xi_0$.  Thus, the use of a ``snow-plough'' approximation is
not unreasonable.  This is supported observationally by the fact that
the three pulsar bow shocks studied to date, in particular the two
with velocities comparable to that of \rxj\ (\object{PSR B1957+20},
\cite{kh88}; \object{PSR J0437$-$4715}, \cite{bbm+95}), display
nebulae with sharp boundaries.

Second, the ambient gas has to be partially neutral in order to see
Balmer line emission. The inferred Hydrogen number density of
$\sim\!1\un{cm^{-3}}$ is not an unreasonable density (in order) for
the Warm Neutral Medium (WNM) or the Warm Ionised Medium (WIM). The
two phases taken together are expected to occupy 50\% of interstellar
space locally; see \citet{kh88b} and \citet{dl90} for general reviews
of the conditions in the interstellar medium.  The ionisation in the
WNM is low but the WIM is partially ionised, $\xi_0\sim 0.5$ (see
\citealt{rtk+95,rl00}).

Independent of the nature of the ambient gas, the bow-shock model
requires that the incoming gas not be fully ionised by the extreme
ultra-violet radiation from \rxj.  In the next section, we treat the
case of a pure ionisation nebula.  We find that the ambient gas is
becomes half-ionised at an angular offset, in the forward direction of
only $\sim\!0\farcs2 f_{\rm bb}$ (Eq.~\ref{eq:theta0}) where $f_{\rm
bb}$ is the excess of ionising flux (in the extreme ultraviolet band)
over that expected from blackbody fits to the X-ray and optical data
of \rxj. Thus we would have to abandon the pulsar bow shock model if
$f_{\rm bb}\simgt4$.

Third, in our estimates, we implicitly assumed the bow-shock was thin
compared to the distance to the neutron star.  \cite{grs+01} have
computed the ionisation structure of Balmer-dominated shocks for
$n_{\rm H^0}=1\un{cm^{-3}}$ and $\xi_0=0.5$.  For a
$250\un{km}\un{s^{-1}}$ shock, they find that the time scale on which
an \ion{H}{i} is ionised by electrons is $0.2\un{yr}$, and that the
time scale on which proton charge exchange becomes ineffective is
about $\sim\!0.4\un{yr}$.  These time scales are expected to depend
only weakly on velocity and to be inversely proportional to density.
The corresponding length scales are $l\simeq10^{14}v_{\rm
ns,100}\un{cm}$, which is smaller than $r_{\rm
a}\simeq10^{15}d_{60}\un{cm}$.  Thus, we conclude that the assumption
of a thin bow shock is reasonable.

Finally, we verify whether the assumption of case A was reasonable.
For the Lyman lines, the line-centre cross-section is given by
(\citealt{rl79})
\begin{eqnarray}
\alpha_{\nu_0,1,n} &=& \frac{\pi e^2}{m_{\rm e}c}f_{1,n}
   \frac{c}{\nu_0}\sqrt{\frac{m_{\rm H}}{2\pi kT}}\nonumber\\
   &=& 1.16\times10^{-14}\un{cm^2}\;
       \lambda_{0,\hbox{\scriptsize\angstrom}} f_{1,n} T_{\rm K}^{-1/2},
\label{eq:taulbeta}
\end{eqnarray}
where $f_{1,n}$ is the oscillator strength (for Ly$\beta$ through
Ly$\epsilon$, $f_{1,2\ldots5}=0.4162$, 0.07910, 0.002899, and
0.001394). The line-centre optical depth in the Ly$\beta$ line is thus
$\sim\!2\times\alpha_{\nu_0,1,1}n_{\rm H^0}l\simeq0.8$; here the
multiplicative factor of 2 is a crude approximation to account for the
bow shock geometry.  Since in every absorption, there is a probability
of only 0.118 that Ly$\beta$ is converted to H$\alpha$ (plus
two-photon continuum), we conclude that our assumption of case A is
reasonable.

\section{An ionisation nebula?}\label{sec:ionisation}

\rxj\ is a source of ionising photons and, as with hot white dwarfs,
should have an ionised region around it.  Such a region would produce
H$\alpha$ emission, and would be deformed because of the proper motion
of the neutron star.  Indeed, \citet{bwm95} showed that isolated
neutron stars accreting from the interstellar medium should have
associated H$\alpha$ nebulae with cometary shape.  

To see whether ionisation could be important for the nebula around
\rxj, we will first make analytical estimates in \Sref{ion:analytic},
following the analysis of \citet{bwm95}.  We will find that ionisation
could lead to a nebula with properties in qualitative agreement with
the observations.  The inferred conditions in the nebula, however, are
rather different from those encountered in ``normal'' ionisation
nebulae; we discuss this briefly in \Sref{ion:normal}.  Next, in
\Sref{ion:model}, we present a detailed simulation with which we
attempt to reproduce the observations quantitatively.

\subsection{Analytical estimates}\label{sec:ion:analytic}

In general, the properties of ionisation nebulae are determined by the
balance between ionisation, recombination, heating and cooling.  In
order to determine the relative importance of these processes, we
first compare their characteristic timescales with the crossing time,
i.e., the time in which the neutron star with velocity $v_{\rm ns}$
moves a distance equal to the distance $r_{\rm a}$ to the apex of the
nebula,
\begin{equation}
t_{\rm cross} = \frac{r_{\rm a}}{v_{\rm ns}}
              = \frac{\theta_{\rm a}}{\mu_{\rm ns}}
              = 3\un{yr}.
\end{equation}
For our analytical estimates here and below we take $r_{\rm
a}=d\theta_{\rm a}\simeq9\times10^{14}d_{60}\un{cm}$ and $v_{\rm
ns}=d\mu_{\rm ns}\simeq100d_{60}\un{km}{s^{-1}}$, i.e., we ignore the
effects of changes in inclination and the difference between observed
proper motion and motion relative to the local interstellar medium
discussed in~\Sref{orientation}.  We will also assume pure Hydrogen
gas.  We will include Helium and use the correct geometry in the
numerical model described in \Sref{ion:model}.

\subsubsection{Ionisation and recombination timescales}
\label{sec:ion:ionisation}

The typical time for a neutral particle at the apex of the bow shock
to be ionised by the emission from the neutron star is
\begin{equation}
t_{\rm ion}
  = \left(\frac{N_{\rm X}}{4\pi r_{\rm a}^2}\alpha_{\rm ion}\right)^{-1}
  = \left(\frac{n_{\rm X}\alpha_{\rm ion}}{\theta_{\rm a}^2}\right)^{-1},
\label{eq:tiondef}
\end{equation}
where $N_{\rm X}$ is the rate at which ionising photons are emitted,
$n_{\rm X}=N_{\rm X}/4\pi d^2$ the rate at which these would arrive at
Earth in the absence of interstellar absorption, and $\alpha_{\rm
ion}$ the effective cross section for ionisation of Hydrogen.  The
latter is given by
\begin{equation}
\alpha_{\rm ion} 
  = \frac{\int_{\nu_{\rm ion}}^\infty \alpha_\nu (S_\nu/h\nu)\,\d\nu}
         {\int_{\nu_{\rm ion}}^\infty (S_\nu/h\nu)\,\d\nu}
  \simeq 1.4\times10^{-19}\un{cm^2}\;T_{\rm ns,50}^{-1.67};
\label{eq:alphaioneff}
\end{equation}
here $S_\nu$ is the surface emission, which for the numerical estimate
is assumed to be a black-body with temperature $kT_{\rm eff}=50T_{\rm
ns,50}\un{eV}$.  For this cross section, the effective optical depth
$\tau=n_{\rm H^0}\alpha_{\rm ion}r_{\rm a}\simeq
1.3\times10^{-4}n_{\rm H^0\!,1}$ is very small (here $n_{\rm
H^0}=1n_{\rm H^0\!,1}\un{cm^{-3}}$ is the neutral Hydrogen number
density).  Even the optical depth at the Lyman edge is small:
$\alpha_{\rm edge}=6.3\times10^{-18}\un{cm^2}$ and hence $\tau_{\rm
edge}\simeq6\times10^{-3}n_{\rm H^0\!,1}$.

We estimate $n_{\rm X}$ by scaling to the observed, de-reddened
optical flux, and assuming that the emitted spectrum resembles that of
a black body, i.e., 
\begin{equation}
n_{\rm X}=n_{\rm opt}
\frac{\int_{\nu_{\rm ion}}^\infty (S_\nu/h\nu)\,\d\nu}
{\int_{\rm opt} (S_\nu/h\nu)\,\d\nu}
\simeq 0.4\un{s^{-1}}\un{cm^{-2}}\;f_{\rm bb}T^{2.0}_{\rm ns, 50},
\label{eq:nx}
\end{equation}
where we used the unabsorbed photon rate inferred in Paper~I from the
optical and ultra-violet photometry of
$8.5\times10^{-8}\un{s^{-1}}\un{cm^{-2}}\un{\AA^{-1}}$ at 5000\un{\AA} 
and where $f_{\rm bb}$ is a factor which takes into account the
extent to which the emitted spectrum deviates from that of a black
body.


With the above estimates, we find
\begin{equation}
t_{\rm ion} \simeq 13\un{yr}\;f_{\rm bb}^{-1}T_{\rm ns,50}^{-0.33}.
\label{eq:tionest}
\end{equation}

The typical time it takes for a proton to recombine is
\begin{equation}
t_{\rm rec} = \left(\alpha_{\rm rec} n_{\rm e}\right)^{-1}
\simeq 10^5\un{yr}\;n_{\rm e,1}^{-1},
\end{equation}
where $\alpha_{\rm rec}$ is the recombination cross section and
$n_{\rm e}=1n_{\rm e,1}\un{cm^{-3}}$ the electron number density.  For
the estimate, we used $\alpha_{\rm
rec}\simeq4\,10^{-13}\un{cm^{-3}}\un{s^{-1}}$, which is for case~A at
$10^4\un{K}$ (\citealt{ost89}).  Case~A is appropriate here, since we
found above that the Lyman continuum is optically thin and thus
recombination directly to the ground state does not lead to local
ionisation of another atom (as is assumed in Case~B, for which the
recombination rate is $\sim\!40\%$ lower).  We will find below that
the temperature is likely substantially higher than $10^4\un{K}$, but
this will only lead to lower recombination rates.

Clearly, for any reasonable density, the recombination time scale is
far longer than the crossing time.  The ionisation time scale, however,
is comparable to the crossing time.  To match the two, $f_{\rm bb}$
will have to be somewhat larger than unity, but such a deviation may
simply reflect the difference between a real neutron star atmosphere
and a black body spectrum.  Thus, the size of the nebula is consistent
with it being due to ionisation.  The tail of ionised matter will have
length of $\mu_{\rm ns}t_{\rm rec}\simeq10\degr$.  This is far
longer than the observed length, but we will see in
\Sref{ion:heating}, where we consider thermal balance and the relevant
emission processes, that this difference can be understood.  First,
however, we calculate the ionisation structure of the head of the
ionisation nebula.

\subsubsection{Structure of the ionisation nebula}
\label{sec:ion:structure}

In general, the fraction of ionised particles $\xi_+=n_{\rm
H}^+/n_{\rm H}$ changes as a function of time as
\begin{equation}
\frac{{\rm d} \xi_+}{{\rm d} t} =
(1-\xi_+)\alpha_{\rm ion}\frac{N_{\rm X}{\rm e}^{-\tau(r)}}{4\pi r^2} 
-\xi_+n_{\rm e}\alpha_{\rm rec}.
\label{eq:dxidt}
\end{equation}
Given our estimates above, both optical depth effects and
recombination can be ignored.  For this case, the ionisation structure
can be solved analytically.  In a cylindrical coordinate system
$(p,z)$ with the neutron star at the origin and gas moving by in the
$+z$ direction, the solution is
\begin{equation}
\xi_+ = 1-\exp\left(-\frac{z_0}{p}\arctan\frac{p}{-z}\right),
\label{eq:xi}
\end{equation}
where the length scale $z_0$ is given by
\begin{equation}
z_0=\frac{\alpha_{\rm ion}^{\rm eff} N_{\rm X}}{4\pi v_{\rm ns}},
\label{eq:z0}
\end{equation}
and where $\arctan(p/{-z})$ -- taken to be in the range $[0,\pi]$ --
measures the angle with respect to the direction of the proper motion.
From \Eref{xi}, one sees that in a parcel of interstellar matter at
given impact parameter $p$, $\xi_+$ increases slowly at first, then
more rapidly as it passes the neutron star, and finally slowly
approaches its asymptotic value of $1-\e^{-\pi z_0/p}$.  For large
$p$, $\xi_+\propto1/p$, as expected when integrating over a $1/r^2$
distribution.

The angular size corresponding to the length scale $z_0$ is
\begin{equation}
\theta_0
  = \frac{z_0}{d}
  = \frac{\alpha_{\rm ion}^{\rm eff} n_{\rm X}}{\mu_{\rm ns}}
  = 0\farcs2\;f_{\rm bb} T_{\rm ns,50}^{0.33}
\label{eq:theta0}
\end{equation}
Like above, we find that to match the observed size, a somewhat larger
ionising photon rate appears required than expected based on the
observed optical and X-ray fluxes, i.e., $f_{\rm bb}>1$.  We will
return to this in \Sref{ion:model}, where we describe our more
detailed modelling.

\subsubsection{Thermal balance and emission process}
\label{sec:ion:heating}

We first ignore all cooling and calculate the temperature due to
heating by the photo-ionisation.  Using a typical energy
$\bar\epsilon$ of the ionising photons of
\begin{equation}
\bar\epsilon
  = \frac{\int_{\nu_{\rm ion}}^\infty h\nu\alpha_\nu (S_\nu/h\nu)\,\d\nu}
         {\int_{\nu_{\rm ion}}^\infty \alpha_\nu (S_\nu/h\nu)\,\d\nu}
  \simeq 31.2\un{eV}\;T_{\rm ns,50}^{0.25},
\label{eq:epsilon}
\end{equation}
we find for a pure Hydrogen gas
\begin{equation}
T = \frac{2}{3k}\frac{\xi_+}{1+\xi_+}(\bar\epsilon-\chi_{1,\infty})
  \simeq 7\times10^4\un{K}\,T_{\rm ns,50}^{0.45}\frac{2\xi_+}{1+\xi_+},
\label{eq:t}
\end{equation}
where $\chi_{1,\infty}=13.6\un{eV}$ is the ionisation potential of
hydrogen.

The high inferred temperature of $\sim\!5\times10^4\un{K}$ for
$\xi_+=0.5$ has an important consequence, viz., that collisional
excitation and ionisation of Hydrogen are important.  These processes
will dominate the cooling.  Furthermore, the H$\alpha$ emission will
be dominated by radiative decay of collisionally excited neutrals.
Below, we first estimate the density required to produce the observed
H$\alpha$ brightness.  Next, we use this to estimate the cooling
time scale and check consistency.

The rate for collisional excitation to the $n=3$ level is $n_{\rm e}
n_{\rm H^0} q_{1,3} = \xi_+(1-\xi_+)n_{\rm H}^2 q_{1,3}$, with the rate
coefficient $q_{1,3}$ given by (see \citealt{ost89})
\begin{equation}
q_{1,3} = \Upsilon_{1,3}\frac{2\sqrt{\pi}\alpha c a_0^2}{\omega_1}
          \sqrt{\frac{\chi_{1,\infty}}{kT}}\;
           {\rm e}^{-\chi_{1,3}/kT},
\label{eq:ralpha}
\end{equation}
where $\Upsilon_{1,3}$ is the sum of collision strengths to the 3s,
3p, and 3d states, $\omega_1=2$ the statistical weight of the ground
state, $\chi_{1,3}=12.1\un{eV}$ the excitation potential, and
$2\sqrt{\pi}\alpha ca_0^2=2.1716\times10^{-8}\un{cm^3}\un{s^{-1}}$.
The collision strength varies slowly, from $\Upsilon_{1,3}=0.25$ at
$10^4\un{K}$ to 0.43 at $7\times10^4\un{K}$ (\citealt{abbs00}).
Inserting numbers, one finds that the largest volume emission rates
are produced where $\xi_+\simeq0.7$.  In consequence, one expects that
the emission will peak at some distance from the neutron star, and
that the ionisation nebula will have a somewhat hollow appearance.

In order to estimate the hydrogen number density required to reproduce
the observed brightness, we will assume case~A, i.e., the medium is
optically thin to Ly$\beta$ photons, so that of the excitations to the
3p state only 11.8\% lead to emission of an H$\alpha$ photon (for 3s
and 3d, the transition to the ground state is forbidden, so all
excitations lead to H$\alpha$ emission).  We label the appropriately
corrected excitation rate as $q_\alpha^{\rm A}$.  Furthermore, we will
ignore excitations into higher levels which lead to H$\alpha$ photons.
We consider the total H$\alpha$ photon rate integrated over the slit
(with width $w_{\rm slit}=1\arcsec$) perpendicular to the proper
motion.  The expected rate is given by
\begin{eqnarray}
n_{\alpha,\perp}
  &\simeq& \frac{\pi r_\perp^2 dw_{\rm slit} n_{\rm H^0} n_{\rm e}
     q_\alpha^{\rm A}}{4\pi d^2}\nonumber\\
  &=& \frac{1}{4}\theta_\perp^2 dw_{\rm slit} \xi_+(1-\xi_+)n_{\rm H}^2 
      q_\alpha^{\rm A}\nonumber\\
  &\simeq& 1.2\times10^{-6}\un{s^{-1}}\un{cm^{-2}}\;d_{60}n_{\rm H,1}^2,
\label{eq:nalpha}
\end{eqnarray}
where for the numerical estimate we used $\xi_+=0.5$ and
$q_\alpha^{\rm A}\simeq2.2\times10^{-10}\un{cm^3}\un{s^{-1}}$
(appropriate for $T=5\times10^4\un{K}$).  To match the observed photon
rate of $1.7\times10^{-5}\un{s^{-1}}\un{cm^{-2}}$ along the slit, thus
requires
\begin{equation}
n_{\rm H} \simeq 4\un{cm^{-3}}\, d_{60}^{-1/2}.
\label{eq:nh}
\end{equation}
This is somewhat denser than the typical density of the interstellar
medium or the mean density of $N_{\rm H}/d\simeq1\un{cm^{-3}}$ along
the line of sight, but not by a large factor.  Hence, the observed
brightness can be reproduced without recourse to extreme assumptions
about the interstellar medium in which the source moves.

With the above estimate of the density, we can calculate the cooling
time scale and verify {\em a posteriori} that cooling is not important
near the head of the nebula.  As mentioned, at the high inferred
temperatures, the dominant cooling processes are collisional
excitation and ionisation of Hydrogen.  At $T\simeq5\times10^4\un{K}$,
collisional excitation of the $n=2$ level is most important, with a
rate coefficient $q_{1,2}\simeq2\times10^{-9}\un{cm^3}\un{s^{-1}}$
(\Eref{ralpha}, $\chi_{1,2}=10.2\un{eV}$, $\Upsilon_{1,2}=1.4$;
\citealt{abbs00}).  Hence, the cooling time scale $t_{\rm cool}$ is
roughly
\begin{eqnarray}
t_{\rm cool}
  &=& \frac{(n_{\rm H}+n_{\rm e})\frac{3}{2}kT}
           {n_{\rm H^0}n_{\rm e}q_{1,2}\chi_{1,2}}
   =  \frac{\bar\epsilon-\chi_{1,\infty}}
           {(1-\xi_+)n_{\rm H}q_{1,2}\chi_{1,2}}\nonumber\\
&\simeq& 14\un{yr}\;n_{\rm H,4}^{-1}.
\label{eq:tcool}
\end{eqnarray}
This time scale is a little longer than the crossing time, which
confirms that temperatures as high as those inferred above can be
reached.  The time scale is sufficiently short, however, that more
precise estimates will need to take cooling into account.  We will do
this in \Sref{ion:model}.

The relatively short cooling time scale provides the promised answer
to the puzzle posed in \Sref{ion:ionisation}: the tail appears far
shorter than the recombination length of $10\degr$ because the
temperature and hence the emissivity drop rapidly behind the neutron
star.  In principle, the tail will emit recombination radiation, but
this will be at unobservable levels.

With the inferred temperatures, we can also understand why we see
Balmer emission only (\Fref{nebspectrum}).  For Helium, the
temperature is too low for collisional excitation to be important.
This leaves recombination and continuum pumping, neither of which are
efficient enough to produce observable emission in our spectrum (see
also \Sref{ion:verification}).  For
the metals, collisional excitation may well take place, with rate
coefficients comparable to or even larger than those for Hydrogen, but
given their low abundances relative to Hydrogen, the resulting
emission will be unobservable as well.

\subsubsection{Dynamical effects}\label{sec:ion:dynamics}

The heating due to photo-ionisation by the relatively hard photons
from the neutron star will also lead to an overpressure in the gas
near the neutron star relative to that further away.  As a result, the
gas should expand.  We estimate the velocity in the direction
perpendicular to the proper motion from $v_p=a_pt$, where $a_p$ is the
acceleration associated with the pressure gradient perpendicular to
the proper motion and $t$ is the time scale on which the pressure
gradient changes.  The acceleration is given by
\begin{equation}
a_p = -\frac{1}{\rho} \frac{\d P}{\d p}
    = -\frac{2}{3m_{\rm H}}\left(\bar\epsilon-\chi_{1,\infty}\right)
       \frac{\d\xi_+}{\d p}.
\label{eq:dvdt}
\end{equation}
At $z=0$, one has $\d\xi_+/\d p=-(p_0/p^2)\e^{-p_0/p}$, where
$p_0\equiv\frac{\pi}{2}z_0$.  The pressure gradient changes due to
expansion and due to the neutron star passing by.  The time scale for
the latter is much faster, and is approximately given by
\begin{equation}
t \simeq \frac{1}{v_{\rm ns}}\left.\frac{\xi_+}{\d\xi_+/\d z}\right|_{z=0}
  = \frac{1}{v_{\rm ns}} \frac{p^2}{z_0}
    \frac{1-\e^{-p_0/p}}{\e^{-p_0/p}}.
\label{eq:tacc}
\end{equation}
Thus, for the velocity one finds
\begin{eqnarray}
v_p &\simeq&\frac{2\left(\bar\epsilon-\chi_{1,\infty}\right)}
                {3m_{\rm H} v_{\rm ns}}
            \frac{p_0}{z_0}\left(1-\e^{-p_0/p}\right)
    \nonumber\\
    &\simeq& 11\un{km}\un{s^{-1}}\;T_{\rm ns,50}^{0.45}\,d_{60}^{-1},
\label{eq:vexp}
\end{eqnarray}
where the numerical value is for $p=p_0$.  Note that for large $p$,
the expansion velocity becomes independent of $p$.  This is confirmed
by our detailed models, but only if cooling is unimportant.

Because of the expansion, behind the neutron star the gas will be more
tenuous and emit fewer H$\alpha$ photons.  This, in turn, leads to a
more hollow appearance of the emission nebula as a whole.  We include
the expansion in our numerical model, which we describe in
\Sref{ion:model}.  First, however, we discuss briefly why the
properties we derive for the nebula differ greatly from those of
``normal'' ionisation nebulae.

\subsection{Intermezzo: comparison with ``normal'' ionisation nebulae}
\label{sec:ion:normal}

Ionisation nebulae observed around hot stars, such as planetary
nebulae and \ion{H}{ii} regions, typically show a wealth of emission
lines of many elements, and have inferred temperatures around
$10^4\un{K}$ (\citealt{ost89}).  In these, the medium is in (rough)
equilibrium: ionisation is balanced by recombination and heating by
cooling.  The emission from Hydrogen is due to (inefficient)
recombination, while that of the metals is due to (efficient)
collisional excitation; hence, the metal lines are strong despite the
low metal abundances.  The emission is strongest where the matter is
densest, and nebulae with roughly uniform density appear filled (e.g.,
\ion{H}{ii} regions).

The situation for \rxj\ is completely different: there is {\em no}
equilibrium; recombination is not relevant at all and cooling only
marginally so.  The reason is that the neutron star emits only few
ionising photons (because it is small) and moves fast.  As a result,
the nebula is ``matter bounded,'' and much smaller than would be
inferred from the Str\"omgren, ``ionisation bounded''
approximation\footnote{The incorrect application of the Str\"omgren
approximation likely led \citet{pwl+01} to their remark that the
nebula around \rxj\ could not be due to ionisation.} (see
\citealt{bwm95}).  The interstellar medium simply does not notice the
few photons passing by until the neutron star is very close, at which
time the photo-ionisation and heating rates increase far more rapidly
with time than can be balanced by recombination and cooling.

In some sense, in considering the conditions in the ionisation nebula
derived here, it may be more fruitful to compare with the conditions
in a bow shock rather than with those in ``normal'' ionisation
nebulae.  Like for the nebula considered here, in the case of a
bow shock, the ambient medium receives a sudden injection of energy, which
cannot be balanced by cooling, and results in a very high temperature
(\Sref{bowshock}).  In both cases, collisional excitation and
ionisation of Hydrogen are highly important, leading to
Balmer-dominated emission spectra.  The main difference is that for
the case of a bow shock, not only energy but also momentum is injected
into the medium.  The absence of the shock associated with the
momentum injection makes the case of the ionisation nebula a much more
tractable one.

\subsection{A detailed model}\label{sec:ion:model}
Our analytic estimates showed that the nebula could be due to
ionisation, without requiring extreme conditions.  To make a more
quantitative comparison, we made a model along the lines of that
described by \citet{bwm95}, except that we include the dynamics.  For
this purpose, we use the hydrodynamics code {\sc zeus-2d}
(\citealt{sn92}).  We changed the code to allow it to keep track of
abundances of separate ions, and wrote a routine that calculates the
photo-ionisation rates, the associated heating, and the cooling due to
collisional processes.  We only include Hydrogen and Helium (at
cosmic abundances, $X=0.73$ and $Y=0.27$); metals are not important
for the ionisation balance, and, as argued above, do not have their
usual role as coolant in our case.

In our calculations, we take into account ionisation of \ion{H}{i},
\ion{He}{i}, and \ion{He}{ii}, as well as the associated heating.  For
each of the ions, we calculate effective cross sections $\alpha_{\rm
ion}$ and energies $\bar\epsilon$ as in \EEref{alphaioneff}
and~\ref{eq:epsilon}, by integrating a black body photon spectrum over
the appropriate cross sections (\ion{H}{i} and \ion{He}{ii}:
\citealt{ost89}; \ion{He}{i}: \citealt{vfky96}).  We verified that
optical depth effects were negligible.  Note that for hard ionising
spectra, Helium is ionised before Hydrogen, and that the resulting
photo-electrons have larger energy (for $kT_{\rm ns}=50\un{eV}$:
$\bar\epsilon-\chi_{1,\infty}\simeq18\un{eV}$ for Hydrogen and
$\sim\!33\un{eV}$ for both \ion{He}{i} and \ion{He}{ii}).  As a
result, temperatures are high already when the fraction of neutral
Hydrogen is still relatively high, and the nebula will be brighter at
large distances than would be expected for pure Hydrogen.

For the cooling, we include collisional ionisation of Hydrogen
(\citealt{sw91}) as well as collisional excitation of Hydrogen from
the ground state to levels 2--5 (\citealt{abbs00}).  We verified that
recombination was unimportant; free-free radiation is also negligible.
We considered secondary ionisations by the fast ionised electrons,
using the values found by \citet{shu79}, but found that it was
unimportant in any region with $\xi_+>0.05$ and hence had only very
slight overall effect.  For the sake of simplicity, therefore, we have
not included it in the results presented here.

In order to compare our results with the observations, we use the
results of the simulation to calculate the collisional excitation
rates from the ground state to all individual levels 2s to 5g of
Hydrogen (\citealt{abbs00}), and then infer emission in particular
transitions using the cascade matrix.  As before, we assume that all
Lyman lines are optically thin (case~A); we will return to this in
\Sref{ion:verification} below.

\begin{figure*}
\includegraphics[width=180mm]{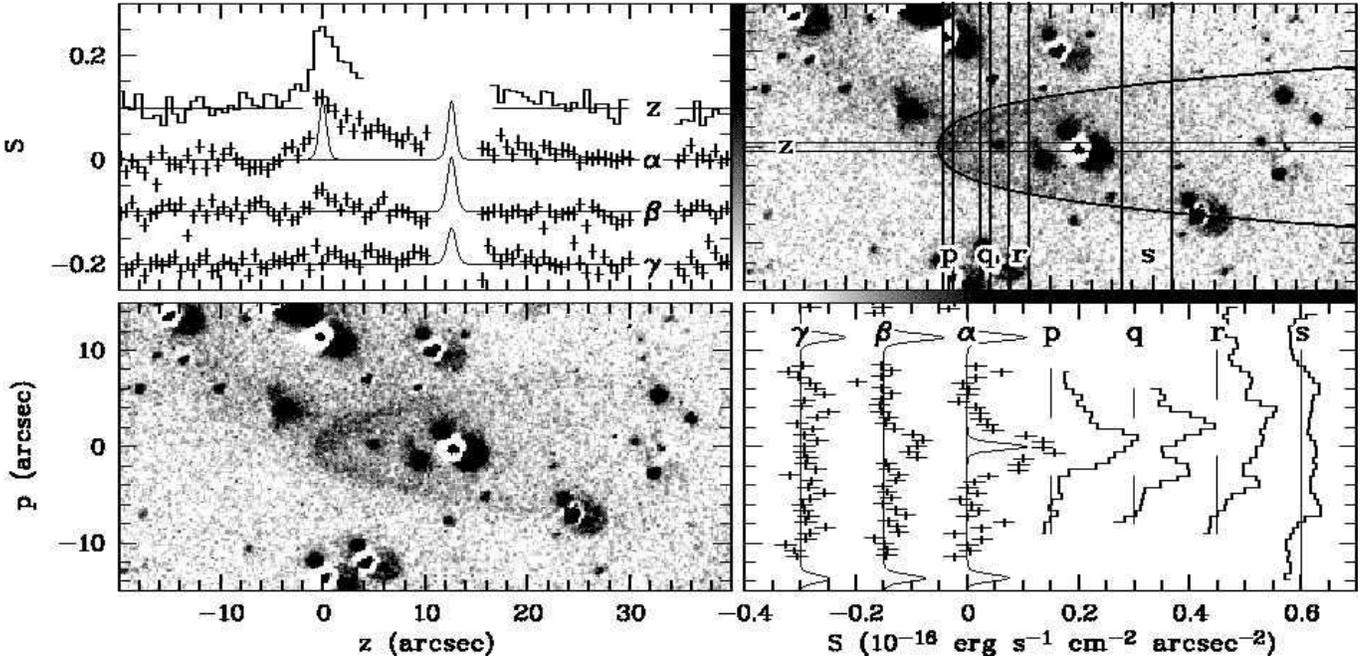}
\caption[]{Model for an ionisation nebula.  The coordinate frame is
cylindrical, with $z$ running along the axis of symmetry and $p$
perpendicular to it.  The neutron star is at the origin and moves to
the left.  The computational domain was between
$-18\times10^{15}<z<36\times10^{15}\un{cm}$ and
$0<p<18\times10^{15}\un{cm}$.  The three left-hand panels show
predicted runs of the fractional abundances of \ion{H}{i},
\ion{He}{i}, and \ion{He}{ii}.  Darker shades indicate lower
fractional abundance; the 20, 40, 60, and 80\% levels are indicated by
contours.  The right three panels show runs of density, temperature,
and H$\alpha$ emissivity.  For the density, darker shade indicates
lower density; the white contours are drawn at 3, 4, 5, 6, and
$7\times10^{-24}\un{g}\un{cm^{-3}}$.  Overlaid are black contours and
arrows indicating the velocity; the contours are drawn at 2, 4, 6, and
$8\un{km}\un{s^{-1}}$.  For the temperature, darker shade indicates
higher temperature; the black contours are drawn at temperature of 1,
1.5, 2, and $2.5\times10^4\un{K}$, and the white ones at 3, 4, and
$5\times10^4\un{K}$.  Finally, for the H$\alpha$ emissivity, darker
shade indicates stronger emissivity; the contours levels are at 1/8,
1/4, 1/2, 1 (black), 2, 4, and
$8\times10^{-10}\un{s^{-1}}\un{cm^{-3}}$ (white).\label{fig:model}}
\end{figure*}

\subsubsection{Model parameters}

Our model requires ten parameters, which describe the conditions in
the undisturbed interstellar medium (1--5; density, temperature, and
fractional \ion{H}{i}, \ion{He}{i}, and \ion{He}{ii} abundances), the
ionising radiation field (6, 7: neutron-star temperature and radius),
the relative velocity between the neutron star and the medium~(8), and
the geometry relative to the observer (9, 10: distance and
inclination).  All but one of these, however, are constrained by
observations.  The distance is around 60\un{pc}, and the inclination
about $60\degr$ assuming an origin in Upper Sco (\Sref{rxj}); the
implied relative velocity is 100\un{km}\un{s^{-1}}
(\Sref{orientation}).  Furthermore, the temperature of the neutron
star is $\sim\!50\un{eV}$ and the effective radius
$7.0\un{km}\,d_{60}(f_{\rm bb}/T_{\rm ns,50})^{1/2}$, where $f_{\rm
bb}$ measures the deviation of the spectrum from that of a black body
(\Eref{nx}) and should be close to unity.  Finally, the interstellar
medium can be in one of four phases, the so-called cold-neutral,
warm-neutral, warm-ionised, and hot-ionised phases (see
\citealt{kh88b} for a review).  Of these, the hot ionised phases
cannot be applicable to our case, since these would not lead to any
emission, while the cold-neutral phase is unlikely because of its
small filling factor.  Furthermore, the warm ionised medium, with
$\xi_+\simeq0.5$ (\Sref{bowshock}) is unlikely, since it typically is
less dense than required.  Thus, we will assume the medium is in the
warm-neutral phase, i.e., it is neutral and has
$T\simeq8\times10^3\un{K}$.  It should have a particle density in the
range 0.1 to $10\un{cm^{-3}}$.

Given the above constraints, the only unknowns are the interstellar
medium density $\rho_{\rm ism}$ and the ionising photon rate 
factor~$f_{\rm bb}$.  As can be seen from the analytical estimates,
these two have rather different effects: the density sets the
brightness of the nebula, while the ionising photon rate sets the
scale.

\subsubsection{Model results}

In \Fref{model}, we show the results of a model calculation with
$\rho_{\rm ism}$ and $f_{\rm bb}$ chosen to roughly reproduce the
observations: $f_{\rm bb}=1.7$ and $\rho_{\rm
ism}=7.7\times10^{-24}\un{g}\un{cm^{-3}}$ (i.e., $n_{\rm
H}=3.4\un{cm^{-3}}$ and $n_{\rm He}=0.31\un{cm^{-3}}$).  The other
parameters are set to the values inferred above: $n_{\rm H^0}/n_{\rm
H}=1$, $n_{\rm He^0}/n_{\rm He}=1$, $n_{\rm He^+}/n_{\rm He}=0$,
$kT_{\rm ns}=50\un{ev}$, $v_{\rm rel}=100\un{km}\un{s^{-1}}$.  The
velocity is appropriate for $i=60\degr$ and $d=60\un{pc}$
(\Sref{orientation}), which are the values used for the comparison
with the observations (\Fref{modelcomp}).

\begin{figure*}
\includegraphics[width=180mm]{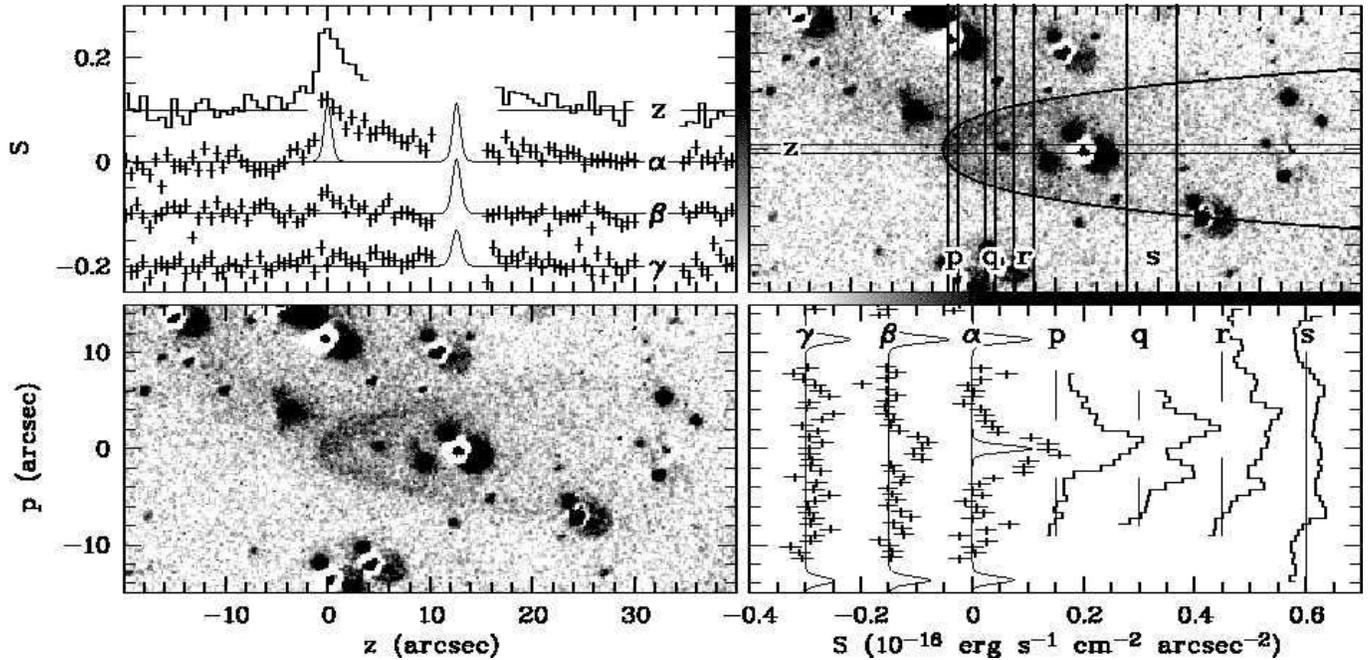}
\caption[]{Comparison of the observations with the predicted Balmer
emission for an ionisation nebula.  As \Fref{nebula}, but with the
predicted H$\alpha$ image shown in the top-right panel, and with
modelled cross cuts overdrawn.  The model reproduces the head of the
nebula quite well, but underpredicts the brightness in the tail.  This
may be because in the tail the assumption of case~A for the line
emission is inappropriate (\Sref{ion:verification}).\label{fig:modelcomp}}
\end{figure*}

In the panels with the fractional abundances in \Fref{model}, one sees
that these follow the pattern expected from the analytic solution of
the structure of the nebula (\Eref{xi}).  Only behind the neutron star
there are deviations, due to the expansion of the nebula.  The
temperature also follows the predictions ahead of the neutron star,
but behind it the compression wave caused by the expansion is visible.
Due to the expansion, the density is reduced by roughly a factor four
far behind the neutron star.  The velocities associated with the
expansion are predominantly perpendicular to the axis of symmetry, as
expected.  The maximum radially outward velocity is
$6.5\un{km}\un{s^{-1}}$ at $2\times10^{15}\un{cm}$ behind the neutron
star.  At large distances, this reduces to $3\un{km}\un{s^{-1}}$ (the
reduction is due to cooling; without cooling, the maximum outward
velocity is about $10\un{km}\un{s^{-1}}$ and remains this high due to
continued photo-ionisation).  No shocks form, since the velocity
remains well below the sound speed everywhere: $(v/c_{\rm s})_{\rm
max}=0.3$.

The predicted H$\alpha$, H$\beta$ and H$\gamma$ fluxes are compared
with the observations in \Fref{modelcomp}.  As one can see, the
agreement near the head of the nebula is very good, but near the tail
the brightness is underpredicted.  

\subsubsection{Fiddling with the parameters}

To see how sensitive our results are to the choice of parameters, we
also ran models for different choices, each time allowing ourselves
the freedom to adjust $\rho_{\rm ism}$ and $f_{\rm bb}$ such that the
match to the observations was as good as possible.  We treat the
different parameters in turn.

First, we consider the properties of the interstellar medium.  If the
undisturbed temperature were lower, all temperatures would be somewhat
lower and hence the emission less efficient.  However, a small change
in $f_{\rm bb}$ will compensate for this: for a medium temperature of
100\un{K}, $f_{\rm bb}=2$ will lead to a model that is hard to
distinguish from our base model.  The medium may also not be
completely neutral, i.e., $\xi_0=n_{\rm H^0}/n_{\rm H}<1$.  This again
does not lead to major differences, as long as one chooses a density
larger by a factor $\sim\!1/\xi_0$, i.e., keeps $n_{\rm H^0}$
approximately constant.  Even for $\xi_0=0.5$, the results are not
much different, except that the tail of the nebula becomes slightly
narrower because of the ``dead mass'' that has to be dragged along in
the expansion.

Second, we varied the temperature of the neutron star.  This will lead
to a change in the number of ionising photons as well as a change in
their mean energy $\bar\epsilon$.  We find, however, that we can
compensate with a suitable change in $f_{\rm bb}$: one requires
$f_{\rm bb}\simeq2.1$ and 1.4 for temperatures $kT_{\rm ns}=40$ and
60\un{eV}, respectively.  The results look very similar, except that
for lower (higher) temperature, the tail is somewhat longer (shorter)
and narrower (wider), and the head is very slightly dimmer (brighter).
Compensation for the change in brightness requires a change in
$\rho_{\rm ism}$ of $\sim\!5\%$.

Third, we considered different inclination.  We found that even for
inclinations as low as 30\degr, the main effect is on the width of the
tail, which becomes narrower; the length of the tail is hardly
affected, while the brightness of the head of the nebula changes only
slightly (brighter for lower inclination).  The somewhat
counter-intuitive result for the width of the tail is a consequence of
the fact that the expansion velocity is inversely proportional to
$v_{\rm ns}$ (\Eref{vexp}), and, therewith proportional to $\sin i$
(since $v_{\rm ns}=v_{\rm ns,\perp}/\sin i =d\mu_{\rm ns}/\sin i$).

Finally, we considered the distance.  For larger distance, less dense
interstellar medium suffices to reproduce the brightness of the head,
as expected from \Eref{nh} (e.g., $\rho_{\rm
ism}=5.4\times10^{-24}\un{g}\un{cm^{-3}}$ for $d=100\un{pc}$).  One
also finds that the tail of the nebula becomes narrower with
increasing distance, as a consequence of the higher space velocity of
the neutron star implied by the measured proper motion.  Indeed, the
width decreases roughly inversely proportional with distance; for
$d=100\un{pc}$, it is substantially narrower than observed.

From our attempts, we conclude that the model results are remarkably
insensitive to fiddling with the parameters.  This is good in the
sense that apparently the observed shape of the head of the nebula
follows quite naturally from the model, but bad in the sense that,
when one matches the brightness of the head, just as naturally one
predicts a tail to the nebula that is too faint.  We discuss below
whether this might be due to one of the assumptions underlying our
model being wrong.

\subsubsection{Verifying the assumptions}
\label{sec:ion:verification}

For models that reproduce the brightness of the head of the nebula,
the brightness in the tail is lower than observed.  Thus, in its
present form, the model is incorrect.  This may relate to one of the
assumptions, which perhaps lead us to overestimate the flux near the
head of the nebula, or underestimate the flux in the tail.  It may be
that we miss an important thermal or emission process, but we argued
above that photo-ionisation of Hydrogen and Helium, and collisional
excitation and ionisation of Hydrogen dominate all other processes.

The one assumption we have not yet verified, is that of case~A for the
line emission, i.e., whether the medium is sufficiently optically thin
to Lyman lines that, e.g., any Ly$\beta$ line produced by collisional
excitation escapes the nebula before being converted in an H$\alpha$
photon (plus two photons in the 2s-1s continuum).  With temperatures
and densities from our model in hand, we can verify whether case~A
actually holds.  

For the Lyman lines, the line-centre cross-section is given in
Eq.~\ref{eq:taulbeta}.  At the position of maximum brightness in the
nebula, $4.5\times10^{14}\un{cm}$ ($0\farcs5$) in front of the neutron
star, our model has $n_{\rm H^0}=1.4\un{cm^{-3}}$ and
$T=5.2\times10^4\un{K}$.  Thus, over the size of the head of the
nebula ($1\arcsec$ or $\sim\!10^{15}\un{cm}$), the line-centre optical
depth for Ly$\beta$ photons is $\tau_{\rm Ly\beta,0}=5.8$.  Averaged
over the Doppler profile, the probability that the photon escapes can
be estimated by $p_{\rm esc}\simeq1.72/(\tau_0+1.72)=0.23$ (here, a
spherical geometry is assumed; \citealt{ost89}, p.~101).  Since every
time a Ly$\beta$ photon is absorbed, there is a probability of only
$p_{\rm conv,1}=0.118$ that it is converted to H$\alpha$ (instead of
being scattered), the total probability that an H$\alpha$ photon is
produced is $p_{\rm conv,tot}\simeq p_{\rm conv,1}/(1-(1-p_{\rm
conv,1})(1-p_{\rm esc}))=0.37$.  This is still substantially smaller
than unity and shows that case~A at least is a better choice than
case~B.  Some of the photons may still be absorbed close to the head
of the nebula, but many may also escape altogether, since in the hot
central regions the Doppler profile is substantially broader than in
the cooler surroundings.  Obviously, for Ly$\gamma$ and Ly$\delta$,
the optical depths are lower and hence the case~A assumption is
better. 

The situation is somewhat different, however, in the tail of the
nebula, where the neutral Hydrogen density is higher and the
temperature lower.  For instance, at $z=1.8\times10^{16}\un{cm}$
(20\arcsec\ behind the neutron star; slit~s), maximum emission is at
$p=5.0\times10^{15}\un{cm}$ ($5\farcs6$ away from the nebular axis),
and at this position one has $n_{\rm H^0}=2.7\un{cm^{-3}}$ and
$T=2.0\times10^4\un{K}$.  Over the relevant length scale at this
position, $\sim\!3\times10^{15}\un{cm}$, the optical depth is
$\tau_{\rm Ly\beta,0}=54$.  Hence, $p_{\rm esc}=0.03$ and $p_{\rm
conv,tot}=0.8$.

From the above, we see that case~A is not quite appropriate.  If one
assumes case~A, therefore, the emissivities are underestimated, and
therewith the required density overestimated.  For the head of the
nebula, where the gas is hot and highly ionised, the effect will not
be very large, but in the tail, where the gas is relatively cool and
less ionised, case~A is clearly inappropriate.  To treat this
correctly, one needs to do full radiation transport, which is beyond
the scope of the present paper.  To get an idea of the effect,
however, we can make the opposite assumption for the tail, namely that
the medium is optically thick to Lyman lines everywhere (i.e.,
case~B).  For this case, the emission in the tail would be increased
by a factor $\sim\!2$, which brings it into much closer agreement with
the observations.  We intend to pursue this in more detail in a future
publication.

One other effect we have ignored so far, is continuum pumping by the
ionising source (\citealt{fer99}), in which neutrals absorb continuum
photons at Lyman transitions and emit Balmer, Paschen, and other
higher-level lines.  For H$\alpha$, we estimate a maximum photon rate
of $10^{-6}\un{s^{-1}}\un{cm^{-2}}$ from this process; for this
estimate, we calculated the neutron star flux at Ly$\beta$ from the
observed optical flux, and assumed that all photons within a thermal
line width of 0.1\un{\AA} (appropriate for $T=5\times10^4\un{K}$) were
absorbed and converted to H$\alpha$ (i.e., case~B).  Even this maximum
rate is well below that produced by collisional excitation near the
head of the nebula.  For the Helium lines, however, continuum pumping
will be important, as the nebular emission is due to recombination and
therefore very weak.  Given the rather large size of the region in
which Helium is ionised, however, the emission will be rather diffuse,
and hence it is no surprise we do not detect it.

\subsubsection{Model inferences}

We found that there were two main free parameters in our model, the
interstellar medium density $\rho_{\rm ism}$ and the factor $f_{\rm
bb}$, which measures the deviation of the spectrum from a black-body.
For the density, the value depends on the fraction $\xi_0=n_{\rm
H^0}/n_{\rm H}$; for a distance of 60\un{pc}, we found that $\rho_{\rm
ism}\simeq7.7\times10^{-24}\xi_0^{-1}\un{g}\un{cm^{-3}}$.  Likely, the
real value is somewhat lower, since in assuming case~A we ignored
conversion of Ly$\beta$ to H$\alpha$ photons, which we found was not
quite correct.  Full radiative transfer calculations are required to
infer a more precise value.  Pending those, we round down the required
Hydrogen number density to $n_{\rm H}\simeq3\xi_0^{-1}\un{cm^{-3}}$.

The factor $f_{\rm bb}$ depends mainly on the neutron-star
temperature.  This is because these two together set the rate and
energy of the ionising photons, the combination of which sets the
total energy input, which is well constrained by the size of the
nebula.  Roughly, one has $E_{\rm in}\propto n_{\rm X}\alpha_{\rm
ion}(\bar\epsilon-\chi)\propto f_{\rm bb}T^{0.78}$.  To keep $E_{\rm
in}$ constant, thus requires $f_{\rm bb}\propto T^{-0.78}$.  Our model
results confirm this expectation; we find that roughly $f_{\rm
bb}=1.7T_{50}^{-1}$.

\section{Ramifications}\label{sec:conclusions}

\rxj\ is the nearest neutron star and the brightest soft X-ray source
(in this category; \citealt{ttz+00}).  As such, it is commanding great
interest, in particular from spectroscopic X-ray missions.  The
available data inform us that \rxj\ is nearby (60\un{pc}) and warm
(50\un{eV}), and that possibly it was formed in a nearby stellar
association about a million years ago.  Beyond this, however, we are
very much in the dark about its nature.  

The source is puzzling for two reasons in particular. First, using
the temperature inferred from the X-ray, ultraviolet and optical
energy distribution, the distance derived from the measured parallax
and observed flux imply a radius of only $\sim\!7\un{km}$. This is too
small for {\em any} equation of state (\citealt{pwl+01}).  Second,
\rxj\ does not show any pulsations.  The lack of pulsations in the
radio could be due to the beam not intersecting our line-of-sight.
The lack of X-ray pulsations (\citealt{pwl+01,bzn+01}), however, is
puzzling, since all well-studied cooling neutron stars do show
significant pulsations.

In this paper, we report the discovery of a cometary nebula around
\rxj, shining in H$\alpha$ and H$\beta$.  We have considered two
models to account for the cometary nebula: a pulsar bow shock model
and a photo-ionisation model.  Both models lead to new, though
different, diagnostics that bear upon the origin of this interesting
object.

We first summarise the diagnostics below (\Sref{constraints}) and then
describe what future observational tests could distinguish between the
two models (\Sref{tests}).  We conclude with a brief discussion about
the puzzling absence of pulsations (\Sref{nopulsations}).

\subsection{Constraints Provided from Observations of the Nebula}
\label{sec:constraints}

In both models, we can show that the pulsar is moving through a low
density medium, $n_{\rm H^0}\simeq0.8\un{cm^{-3}}$ (bow-shock model)
or $n_{\rm H^0}\simeq3\un{cm^{-3}}$ (photo-ionisation model).  Such
densities are typical of the warm neutral medium (WNM) and perhaps
even the warm ionised medium (WIM).  These two phases of the
interstellar medium are expected to pervade about half of interstellar
space.  Thus, it is not unusual to find \rxj\ embedded in one of
these.

The two models do not directly constrain the neutral fraction of the
ambient gas, $\xi_0$. The WNM is expected to be essentially neutral,
while for the WIM one finds $\xi_0\simgt0.5$ (at least in those few
cases where sufficient data exist to allow estimates; see
\Sref{bowshock}).  

Below we discuss additional diagnostics that are peculiar to each of
these two models.

\subsubsection{Pulsar bow shock}

``Balmer-dominated'' nebulae have been seen around a few pulsars.  The
pulsar wind combined with the supersonic motion of the pulsar in an
ambient (partially) neutral medium results in a bow shock.  The Balmer
lines arise from the collisional excitation of ambient neutral atoms
as they drift into the post-shock region.

In the framework of the pulsar bow-shock model, \rxj\ is a young
off-beam pulsar with an energetic pulsar wind powered by rotational
energy loss
$\dot{E}\simeq6\times10^{31}\xi_0^{-1}d_{60}^{3}\un{erg}\un{s^{-1}}$.
Combining this $\dot{E}$ with the inferred age of $\sim\!10^6\un{yr}$,
we estimate the spin period ($P$) and the strength of the magnetic
dipole field ($B$): $P\simeq3\chi^{1/2}\xi_0^{1/2}\un{s}$ and
$B\simeq10^{13}\chi\xi_0^{1/2}\un{G}$; here $\chi=1-(P_0/P)^2$ with
$P_0$ being the spin period at birth.  Since $\chi\leq1$ and
$\xi_0\leq1$, the estimates are upper limits; they are consistent with
values for $P$ and $B$ expected from an ordinary million-year old
pulsar.

How secure are these conclusions? At the speed relevant to this source
-- $100\un{km}\un{s^{-1}}$ -- there currently exist no rigorous
calculations of Balmer-dominated shocks.  In addition, we have not
taken into account the pre-ionisation of the ambient gas ahead of the
neutron star by the radiation from the neutron star and from the
shocked gas, we have not properly modelled the bow shock geometry, and
we have ignored issues of radiative transport.  Thus, the constraints
we derive must be viewed with some caution.  With some effort,
however, the full modelling effort can be undertaken (J.~Raymond,
pers.\ comm.).

\subsubsection{Photo ionisation nebula}

We also investigated the possibility that the nebular emission is
caused by photo-ionisation and heating of the ambient gas by the
extreme-ultraviolet radiation of the neutron star.  Such nebulae were
predicted to exist around rapidly moving, hot neutron stars by
\citet{bwm95}.  They should have cometary shape and be rather
different from normal ionisation nebulae, in that both recombination
and metal-line cooling are irrelevant.

We made analytical estimates which showed that the nebula around \rxj\
could indeed be due to photo-ionisation.  Since the input atomic
physics for the photo-ionisation model is well known and relatively
simple, we proceeded by building a detailed model.  We find that the
model can reproduce the observations in some detail.  There are
discrepancies as well, but we argued that those could be related to
the fact that for the line emission neither case~A nor case~B is
applicable, and hence one has to do full radiative transport.

We concluded that an ionisation nebula could reproduce the
observations, provided the ambient density is about
$3\xi_0^{-1}\un{cm^{-3}}$ and that the extreme-ultraviolet flux (which
is responsible for most of the photo-ionisation) is somewhat higher,
by a factor $\sim\!1.7$, than expected based on the black-body
spectrum that best fits the optical-ultraviolet and X-ray fluxes.
Since some deviations of the spectrum from the blackbody distribution
are expected, and since deviations are indeed seen at X-ray
wavelengths, the required excess at extreme ultra-violet wavelengths
does not seem unreasonable.  Indeed, in the context of this model, the
nebula provides a measurement of the extreme ultraviolet flux,
something which is not possible otherwise because of interstellar
extinction.

Proceeding further, in the photo-ionisation framework (but not in the
bow shock model), it is plausible that the ambient gas will accrete
onto the neutron star (if not stopped by a pulsar wind or flung off by
a rotating magnetic field).  Assuming the Bondi-Hoyle mechanism, we
find an accretion radius of $R_{\rm acc}= 2GM_{\rm ns}/v_{\rm ns}^2=
3.7\times10^{12}d_{60}^{-2}\un{cm}$.  This leads to an accretion rate
of $\dot{M}=\pi R_{\rm acc}^2\rho_{\rm ism} v_{\rm
ns}=3\times10^9\,d_{60}^{-3.5}\un{g}\un{s^{-1}}$ (where we used
$n_{\rm H}=3d_{60}^{-0.5}\un{cm^{-3}}$ and $X=0.73$).  The resulting
accretion power is $\sim\!GM_{\rm ns}\dot{M}/R_{\rm ns}\simeq
6\times10^{29}d_{60}^{-3.5}\un{erg}\un{s^{-1}}$.  In contrast, the
bolometric luminosity of \rxj\ is $\sim\!4\times
10^{31}d_{60}^2\un{erg}\un{s^{-1}}$.

Thus accretion, even if takes place, is energetically insignificant.
\rxj\ is hot not because of accretion.  However, accretion will also
introduce cosmic composition material to the surface of the neutron
star.  If the metals settle out rapidly, as is expected for accretion
at levels too low to contribute significantly to the luminosity
(\citealt{bbr98,bsw92}), this would lead to an atmosphere of almost
pure Hydrogen.  This would be inconsistent with the observations
(\citealt{pwl+01,bzn+01}).  


\subsubsection{A combination?}

If the nebula is indeed formed by photo-ionisation then any $\dot{E}$
associated with rotational energy loss should be small enough that the
standoff radius, $r_{\rm a}$, is less than the observed size of the
nebula.  Adopting the density required by the ionisation model,
$n_{\rm H}\simeq3\un{cm^{-3}}$, this leads to an upper limit,
$\dot{E}\simlt2\times10^{32}d_{60}^{3.5}\un{erg}\un{s^{-1}}$.  As
before, combining knowledge of $\dot{E}$ with the age of $10^6\un{yr}$
inferred from an origin in Upper Scorpius, we derive
$P\simgt2\chi^{1/2}\un{s}$ and $B\simgt7\times10^{12}\chi\,$G.
Unfortunately these latter constraints are weak, since
$0\leq\chi\leq1$.

Conversely, if the factor $f_{\rm bb}$ is significantly larger than
unity, $f_{\rm bb}\simgt4$, then the bow shock model is ruled out.

\subsection{Observational Tests}
\label{sec:tests}

The critical difference between the pulsar bow shock model and the
photo-ionisation model is in the velocity field. In the former, we
expect to see more than half (\citealt{ray91}) of the H$\alpha$
emission to have velocity widths comparable to the shock speed of
$100\un{km}\un{s^{-1}}$ both because of the high post-shock
temperature and because of large bulk motion.  In contrast, in the
photo-ionisation model the thermal velocities are moderate,
$\simlt\!40\un{km}\un{s^{-1}}$, and the bulk motion even smaller,
$\simlt\!10\un{km}\un{s^{-1}}$ (see Fig.~\ref{fig:model}).  Thus the
simplest test is to obtain higher spectral resolution data.  If only
narrow lines are seen, then the photo-ionisation model is vindicated.
With adequate spatial resolution, one can map the velocity field, for
which again the expectations are clearly different in the two models.

A high spatial resolution image of the nebula would be extremely
useful.  First, independent of which model is correct, such an
observation can be used to derive the direction of the velocity vector
of \rxj\ (because of the intrinsic cylindrical symmetry of both
models).  Next, it can be used to see whether the nebula is thin with
a sharp edge, as expected in the bow shock model, or whether its
emission gradually peters out, as expected in the photo-ionisation
model.  A pulsar wind would evacuate gas up to its stand-off radius
and thus the most interesting possible result of a high resolution
image would be the discovery of a cavity around \rxj, contained {\em
within} the nebula.  If so, the photo-ionisation model would be
vindicated and we would still obtain a direct measure of $\dot E$.

\subsection{Lack of X-ray Pulsations}\label{sec:nopulsations}

From the proper motion, it seems likely that \rxj\ originated in the
Upper Sco OB association and that it is a a million-year old cooling
neutron star.  Indeed, for the bow-shock model, the inferred value of
an $\dot{E}$ is typical of a million year old pulsar.  If so, why does
\rxj\ not exhibit pulsations?  A radio beam might not intersect our
line-of-sight, but all well-studied cooling neutron stars show
significant pulsations at X-ray wavelengths.

The pulsation in cooling radiation arises because magnetic fields
influence the conductivity in the crust and thereby lead to
temperature variation at the surface. A hotter thermal component (in
addition to any magnetospheric emission) is seen in energetic pulsars
and usually attributed to a polar cap heated by pulsar activity.

Among the middle-aged pulsars, we note with interest that the one with
the weakest field, \object{PSR B1055$-$52} ($P=197\,$ms, $B\sim
1.1\times 10^{12}\,$G and $\dot E\sim 3\times 10^{34}\,$erg s$^{-1}$),
is also the one whose flux shows weakest modulation, $\simlt\!10\%$
(\citealt{of93}).  Could it be that \rxj\ is a slightly older analogue
of \object{PSR B1055$-$52}?  The pulsed fraction in the cooling
emission for \rxj\ could be reduced further by appealing to an even
weaker magnetic field strength, say to a few $10^{11}\,$G (in the
bow-shock picture, this would imply that $\chi\ll1$, i.e., that the
pulsar was born with a period close to the current one).  If so, it
might have been born with properties like those of the 39.5-ms pulsar
\object{PSR B1951+32}, which has $B\simeq5\times10^{11}\un{G}$ and is
associated with the supernova remnant \object{CTB 80}
(\citealt{kcb+88}).

If the suggestion that \rxj\ has a magnetic field of a few
$10^{11}\un{G}$ is correct, we expect the following: (1) more
sensitive observations will reveal pulsations at the level of a few
percent; (2) $\dot{E}$, when measured, will be below
$2\times10^{32}\un{erg}\un{s^{-1}}$; (3) at high sensitivity, a weak
hot component, $\sim 1\%$ of the total emitted flux,
may be identified.  All these expectations can be tested
with deep {\em Chandra} and {\em XMM} observations.

Most pulsars that we observe, however, appear to be born with magnetic
field strengths upward of $10^{12}\un{G}$.  In this respect, the {\em
a priori} probability seems low that \rxj\ has such a weak field.  It
may be that, after all, \rxj\ did not originate in Upper Sco, and is
old, lost its magnetic field due to accretion, and perhaps was
recently reheated by passing through a dense molecular cloud.  Or
perhaps \rxj\ was just born with a very low magnetic field strength,
or lost its field in a binary with a complicated history.  These
possibilities seem even less plausible.  Thus, if our chain of logic,
which admittedly is long and has many weak links, is correct, \rxj\ is
a middle-aged relatively weakly magnetised pulsar whose beam does not
intersect our line of sight.

It is ironic that the two brightest nearby neutron stars, \rxj\ and
\rxjm, may well represent the extreme ends of the neutron star
magnetic field distribution, one a weak field neutron star and another
a magnetar.

\begin{acknowledgements}
We thank Peter van Hoof for help with finding up-to-date collision
strengths of Hydrogen, Eric van de Swaluw for first model calculations
of heating and expansion, and Omer Blaes, Lars Bildsten, John Raymond
and Parviz Ghavamian for useful discussions.  We thank Drs Stone and
Norman for making their hydrodynamics code ZEUS-2D -- as well as a
clear description -- publicly available.  This research made extensive
use of NASA's Astrophysics Data System Abstract Service and of CDS's
SIMBAD database.  MHvK acknowledges support of a fellowship from the
Royal Netherlands Academy of Science. SRK's research is supported by
NSF and NASA.

\end{acknowledgements}
\bibliographystyle{apj}
\bibliography{rxj1856}

\begin{thebibliography}{49}
\expandafter\ifx\csname natexlab\endcsname\relax\def\natexlab#1{#1}\fi

\bibitem[{{Alard} \& {Lupton}(1998)}]{al98}
{Alard}, C. \& {Lupton}, R.~H. 1998, \apj, 503, 325

\bibitem[{{Anderson} {et~al.}(2000){Anderson}, {Ballance}, {Badnell}, \&
  {Summers}}]{abbs00}
{Anderson}, H., {Ballance}, C.~P., {Badnell}, N.~R., \& {Summers}, H.~P. 2000,
  J.\ Phys.\ B, 33, 1255

\bibitem[{{Bell} {et~al.}(1995){Bell}, {Bailes}, {Manchester}, {Weisberg}, \&
  {Lyne}}]{bbm+95}
{Bell}, J.~F., {Bailes}, M., {Manchester}, R.~N., {Weisberg}, J.~M., \& {Lyne},
  A.~G. 1995, \apjl, 440, L81

\bibitem[{{Bildsten} {et~al.}(1992){Bildsten}, {Salpeter}, \&
  {Wasserman}}]{bsw92}
{Bildsten}, L., {Salpeter}, E.~E., \& {Wasserman}, I. 1992, \apj, 384, 143

\bibitem[{{Blaes} {et~al.}(1995){Blaes}, {Warren}, \& {Madau}}]{bwm95}
{Blaes}, O., {Warren}, O., \& {Madau}, P. 1995, \apj, 454, 370

\bibitem[{{Brown} {et~al.}(1998){Brown}, {Bildsten}, \& {Rutledge}}]{bbr98}
{Brown}, E.~F., {Bildsten}, L., \& {Rutledge}, R.~E. 1998, \apjl, 504, L95

\bibitem[{{Burwitz} {et~al.}(2001){Burwitz}, {Zavlin}, {Neuh{\"a}user},
  {Predehl}, {Tr{\"u}mper}, \& {Brinkman}}]{bzn+01}
{Burwitz}, V., {Zavlin}, V.~E., {Neuh{\"a}user}, R., {Predehl}, P.,
  {Tr{\"u}mper}, J., \& {Brinkman}, A.~C. 2001, \aap, submitted

\bibitem[{{Caraveo} {et~al.}(1996){Caraveo}, {Bignami}, \& {Trumper}}]{cbt96}
{Caraveo}, P.~A., {Bignami}, G.~F., \& {Trumper}, J.~E. 1996, \aapr, 7, 209

\bibitem[{{Chevalier} \& {Raymond}(1978)}]{cr78}
{Chevalier}, R.~A. \& {Raymond}, J.~C. 1978, \apjl, 225, L27

\bibitem[{{Cordes} {et~al.}(1993){Cordes}, {Romani}, \& {Lundgren}}]{crl93}
{Cordes}, J.~M., {Romani}, R.~W., \& {Lundgren}, S.~C. 1993, \nat, 362, 133

\bibitem[{{Cox} \& {Raymond}(1985)}]{cr85}
{Cox}, D.~P. \& {Raymond}, J.~C. 1985, \apj, 298, 651

\bibitem[{{Dehnen} \& {Binney}(1998)}]{db98}
{Dehnen}, W. \& {Binney}, J.~J. 1998, \mnras, 298, 387

\bibitem[{{Dickey} \& {Lockman}(1990)}]{dl90}
{Dickey}, J.~M. \& {Lockman}, F.~J. 1990, \araa, 28, 215

\bibitem[{{Ferland}(1999)}]{fer99}
{Ferland}, G.~J. 1999, \pasp, 111, 1524

\bibitem[{{Gaensler} {et~al.}(2000){Gaensler}, {Stappers}, {Frail}, {Moffett},
  {Johnston}, \& {Chatterjee}}]{gsf+00}
{Gaensler}, B.~M., {Stappers}, B.~W., {Frail}, D.~A., {Moffett}, D.~A.,
  {Johnston}, S., \& {Chatterjee}, S. 2000, \mnras, 318, 58

\bibitem[{{Ghavamian} {et~al.}(2001){Ghavamian}, {Raymond}, {Smith}, \&
  {Hartigan}}]{grs+01}
{Ghavamian}, P., {Raymond}, J., {Smith}, R.~C., \& {Hartigan}, P. 2001, \apj,
  547, 995

\bibitem[{{Haberl} {et~al.}(1997){Haberl}, {Motch}, {Buckley}, {Zickgraf}, \&
  {Pietsch}}]{hmb+97}
{Haberl}, F., {Motch}, C., {Buckley}, D. A.~H., {Zickgraf}, F.-J., \&
  {Pietsch}, W. 1997, \aap, 326, 662

\bibitem[{{Hartigan} {et~al.}(1987){Hartigan}, {Raymond}, \&
  {Hartmann}}]{hrh87}
{Hartigan}, P., {Raymond}, J., \& {Hartmann}, L. 1987, \apj, 316, 323

\bibitem[{{Heyl} \& {Kulkarni}(1998)}]{hk98}
{Heyl}, J.~S. \& {Kulkarni}, S.~R. 1998, \apjl, 506, L61

\bibitem[{{Hoogerwerf} {et~al.}(2001){Hoogerwerf}, {de Bruijne}, \& {de
  Zeeuw}}]{hdbz01}
{Hoogerwerf}, R., {de Bruijne}, J.~H.~J., \& {de Zeeuw}, P.~T. 2001, \aap, 365,
  49

\bibitem[{{Kulkarni} {et~al.}(1988){Kulkarni}, {Clifton}, {Backer}, {Foster},
  \& {Fruchter}}]{kcb+88}
{Kulkarni}, S.~R., {Clifton}, T.~C., {Backer}, D.~C., {Foster}, R.~S., \&
  {Fruchter}, A.~S. 1988, \nat, 331, 50

\bibitem[{{Kulkarni} \& {Heiles}(1988)}]{kh88b}
{Kulkarni}, S.~R. \& {Heiles}, C. 1988, in Galactic and Extragalactic Radio
  Astronomy, ed. G.~L. {Verschuur} \& K.~I. {Kellermann} (Heidelberg,
  Springer), 95--153

\bibitem[{{Kulkarni} \& {Hester}(1988)}]{kh88}
{Kulkarni}, S.~R. \& {Hester}, J.~J. 1988, \nat, 335, 801

\bibitem[{{Kulkarni} \& {van Kerkwijk}(1998)}]{kvk98}
{Kulkarni}, S.~R. \& {van Kerkwijk}, M.~H. 1998, \apjl, 507, L49

\bibitem[{{Lattimer} \& {Prakash}(2001)}]{lp01}
{Lattimer}, J.~M. \& {Prakash}, M. 2001, \apj, 550, 426

\bibitem[{{McKee} \& {Hollenbach}(1980)}]{mh80}
{McKee}, C.~F. \& {Hollenbach}, D.~J. 1980, \araa, 18, 219

\bibitem[{{Motch}(2000)}]{mot00}
{Motch}, C. 2000, astro-ph/0008485

\bibitem[{{Motch} \& {Haberl}(1998)}]{mh98}
{Motch}, C. \& {Haberl}, F. 1998, \aap, 333, L59

\bibitem[{{\"Ogelman} \& {Finley}(1993)}]{of93}
{\"Ogelman}, H. \& {Finley}, J.~P. 1993, \apjl, 413, L31

\bibitem[{{Osterbrock}(1989)}]{ost89}
{Osterbrock}, D.~E. 1989, Astrophysics of gaseous nebulae and active galactic
  nuclei (Mill Valley, CA, University Science Books)

\bibitem[{{Pons} {et~al.}(2001){Pons}, {Walter}, {Lattimer}, {Prakash},
  {Neuh{\"a}user}, \& {An}}]{pwl+01}
{Pons}, J.~A., {Walter}, F.~M., {Lattimer}, J.~M., {Prakash}, M.,
  {Neuh{\"a}user}, R., \& {An}, P. 2001, \apj, submitted (astro-ph/0107404)

\bibitem[{{Predehl} \& {Kulkarni}(1995)}]{pk95}
{Predehl}, P. \& {Kulkarni}, S.~R. 1995, \aap, 294, L29

\bibitem[{{Predehl} \& {Schmitt}(1995)}]{ps95}
{Predehl}, P. \& {Schmitt}, J. H. M.~M. 1995, \aap, 293, 889

\bibitem[{{Raymond}(1991)}]{ray91}
{Raymond}, J.~C. 1991, \pasp, 103, 781

\bibitem[{{Redfield} \& {Linsky}(2000)}]{rl00}
{Redfield}, S. \& {Linsky}, J.~L. 2000, \apj, 534, 825

\bibitem[{{Reynolds} {et~al.}(1995){Reynolds}, {Tufte}, {Kung}, {McCullough},
  \& {Heiles}}]{rtk+95}
{Reynolds}, R.~J., {Tufte}, S.~L., {Kung}, D.~T., {McCullough}, P.~R., \&
  {Heiles}, C. 1995, \apj, 448, 715

\bibitem[{{Rybicki} \& {Lightman}(1979)}]{rl79}
{Rybicki}, G.~B. \& {Lightman}, A.~P. 1979, Radiative processes in astrophysics
  (New York, Wiley-Interscience)

\bibitem[{{Scholz} \& {Walters}(1991)}]{sw91}
{Scholz}, T.~T. \& {Walters}, H. R.~J. 1991, \apj, 380, 302

\bibitem[{{Shull}(1979)}]{shu79}
{Shull}, J.~M. 1979, \apj, 234, 761

\bibitem[{{Shull} \& {McKee}(1979)}]{sm79}
{Shull}, J.~M. \& {McKee}, C.~F. 1979, \apj, 227, 131

\bibitem[{{Stetson}(1987)}]{ste87}
{Stetson}, P.~B. 1987, \pasp, 99, 191

\bibitem[{{Stone} \& {Norman}(1992)}]{sn92}
{Stone}, J.~M. \& {Norman}, M.~L. 1992, \apjs, 80, 753

\bibitem[{{Treves} {et~al.}(2000){Treves}, {Turolla}, {Zane}, \&
  {Colpi}}]{ttz+00}
{Treves}, A., {Turolla}, R., {Zane}, S., \& {Colpi}, M. 2000, \pasp, 112, 297

\bibitem[{{Van Kerwijk} \& {Kulkarni}(2001)}]{vkk01}
{Van Kerwijk}, M.~H. \& {Kulkarni}, S.~R. 2001, {\aap, accepted},
  astro-ph/0106265 (Paper~I)

\bibitem[{{Verner} {et~al.}(1996){Verner}, {Ferland}, {Korista}, \&
  {Yakovlev}}]{vfky96}
{Verner}, D.~A., {Ferland}, G.~J., {Korista}, K.~T., \& {Yakovlev}, D.~G. 1996,
  \apj, 465, 487

\bibitem[{{Walter}(2001)}]{wal01}
{Walter}, F.~M. 2001, \apj, 549, 433

\bibitem[{{Walter} \& {Matthews}(1997)}]{wm97}
{Walter}, F.~M. \& {Matthews}, L.~D. 1997, \nat, 389, 358

\bibitem[{{Walter} {et~al.}(1996){Walter}, {Wolk}, \& {Neuh{\"a}user}}]{wwn96}
{Walter}, F.~M., {Wolk}, S.~J., \& {Neuh{\"a}user}, R. 1996, \nat, 379, 233

\bibitem[{{Wilkin}(1996)}]{wil96}
{Wilkin}, F.~P. 1996, \apjl, 459, L31

\end{thebibliography}
\end{document}